\documentclass[journal, 10pt, twocolumn, a4paper]{IEEEtran}

\usepackage{amssymb}
\usepackage{amsmath}
\usepackage{amsthm}
\usepackage{mathrsfs}
\usepackage{verbatim}
\usepackage{bbm}

\ifCLASSINFOpdf

    \usepackage[pdftex]{graphicx}
    \DeclareGraphicsExtensions{.pdf,.jpeg,.png}
	\usepackage[pdftex, bookmarks = true, colorlinks=false]{hyperref}
\else
	\usepackage[dvips]{graphicx}
	\DeclareGraphicsExtensions{.eps}
	\usepackage[dvips, bookmarks=true, colorlinks=false]{hyperref}
\fi
\ifCLASSOPTIONcompsoc
	\usepackage[caption=false,font=normalsize,labelfont=sf,textfont=sf]{subfig}
\else
	\usepackage[caption=false,font=footnotesize]{subfig}
\fi

\RequirePackage{lineno}

\usepackage{color}
\usepackage[usenames]{xcolor}

\usepackage{arydshln,leftidx,mathtools}
\usepackage{booktabs}

\usepackage{subfig}
\usepackage{multirow}

\usepackage{array}
\newcolumntype{L}[1]{>{\raggedright\let\newline\\\arraybackslash\hspace{0pt}}m{#1}}
\newcolumntype{C}[1]{>{\centering\let\newline\\\arraybackslash\hspace{0pt}}m{#1}}
\newcolumntype{R}[1]{>{\raggedleft\let\newline\\\arraybackslash\hspace{0pt}}m{#1}}

\usepackage{paralist}
\usepackage{enumerate}

\usepackage{enumitem}
\setlist[enumerate]{noitemsep, topsep=0pt}

\makeatletter
\newcommand*{\Rom}[1]{\expandafter\@slowromancap\romannumeral #1@}
\newcommand*{\rom}[1]{\romannumeral #1}
\makeatother

\DeclareMathOperator*{\diag}{diag}

\newtheorem*{theorem*}{Theorem}

\newtheorem*{definition*}{Definition}

\begin{document}

\title{Detector Based Radio Tomographic Imaging}
\author{
	H\"{u}seyin~Yi\u{g}itler, Riku~J\"{a}ntti, Ossi~Kaltiokallio, and Neal~Patwari
    \thanks{ H\"{u}seyin~Yi\u{g}itler, Riku~J\"{a}ntti and Ossi Kaltiokallio are with Aalto University, Department of Communications and
Networking. e-mail: \{name.surname\}@aalto.fi}
	\thanks{ Neal Patwari is with University of Utah, Department of Electrical and Computer Engineering. e-mail: npatwari@ece.utah.edu}
	}
\maketitle

\begin{abstract}
Received signal strength based radio tomographic imaging is a popular device-free indoor localization method which reconstructs the spatial loss field of the environment using measurements from a dense wireless network. Existing methods solve an associated inverse problem using algebraic or compressed sensing reconstruction algorithms. We propose an alternative imaging method that reconstructs spatial field of occupancy using a back-projection based reconstruction algorithm. The introduced system has the following advantages over the other imaging based methods: 
\begin{inparaenum}[i.)]
	\item{significantly lower {computational complexity} such that no floating point multiplication is required;}
	\item{each link's measured data is compressed to a single bit, providing improved scalability;}
	\item{{physically significant} and {repeatable} parameters.}
\end{inparaenum}
The proposed method is validated using measurement data. Results show that the proposed method achieves the above advantages without loss of accuracy compared to the other available methods.
\end{abstract}
\begin{IEEEkeywords}
Device-free localization, radio tomographic imaging, radio wave reflection, signal detection, received signal strength measurements, back-projection reconstruction
\end{IEEEkeywords}

\section{Introduction}
Received signal strength (RSS) measurements of low-cost and low-power radios in dense wireless sensor networks have recently been exploited for gaining situational awareness of the surrounding environment. These deployments can be used for unobtrusive monitoring purposes such as device-free localization (DFL) \cite{Patwari2010} and respiration rate monitoring \cite{Kaltiokallio2014}. The technology enables new applications in for example ambient assisted living \cite{Bocca2013}, residential monitoring \cite{Kaltiokallio2012a}, security and emergency surveillance \cite{Bjorkbom2013} since it does not require the object to cooperate with the system by carrying a tag. It also is an alternative to traditional technologies used for roadside surveillance \cite{Anderson2014} and forest monitoring \cite{Alippi2015}.

\begin{figure}[!t]
\centering
\setlength{\tabcolsep}{0.1cm}
\begin{tabular}{C{0.49\textwidth}}
\subfloat[]{\includegraphics[width=0.49\textwidth]{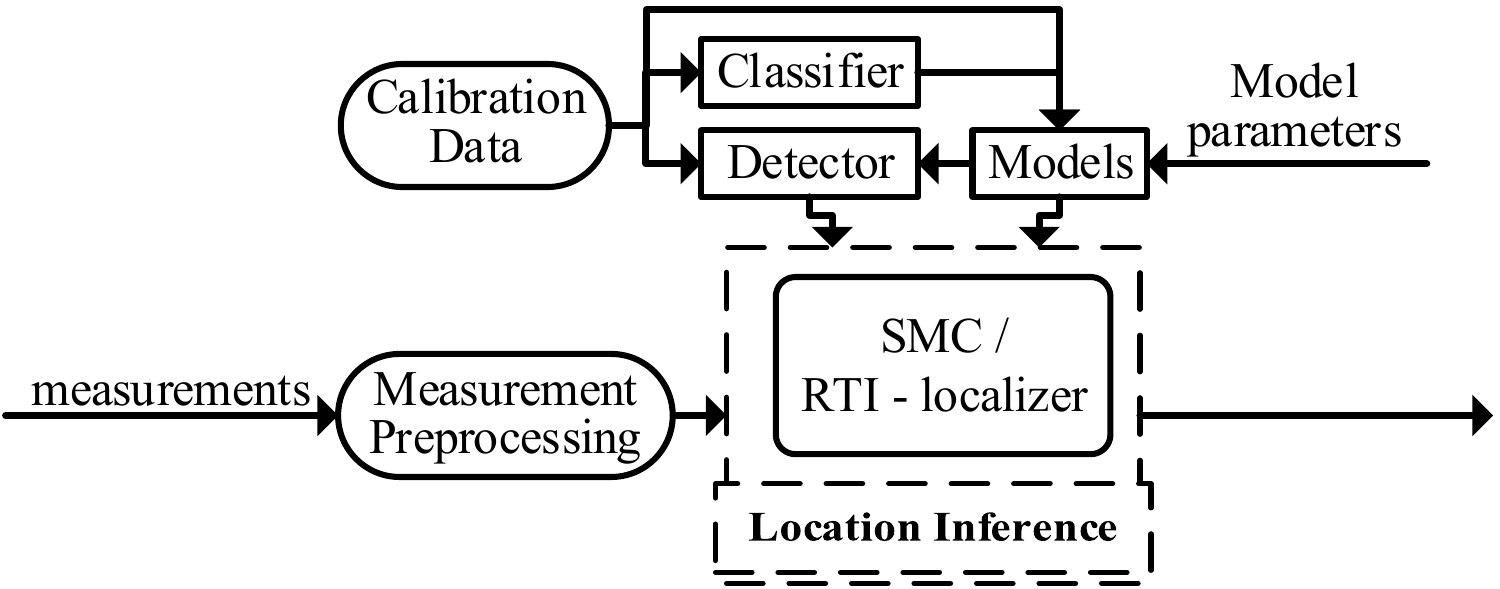}\label{fig:dfl-system}} \\
\subfloat[]{\includegraphics[width=0.49\textwidth]{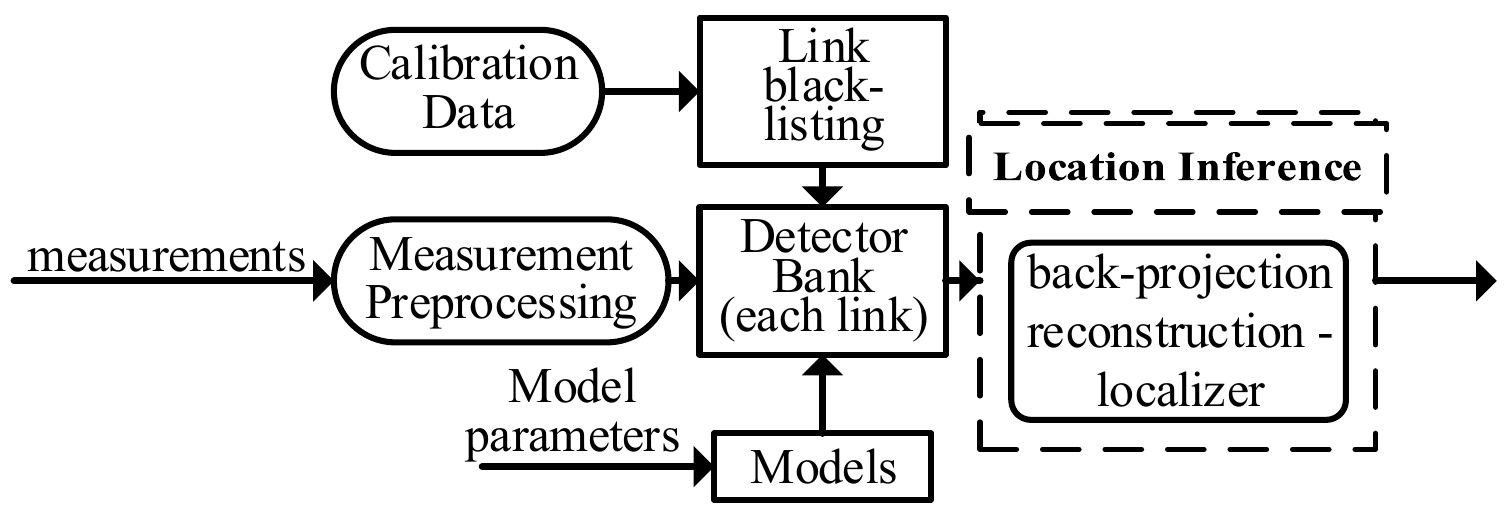}\label{fig:system}}
\end{tabular}
\caption{In (a), typical DFL system. In (b), introduced RTI based DFL system.
}
\end{figure}

RSS-based DFL is built upon the fact that people\footnote{In this paper, we use person and interacting object interchangeably. By these, we refer to any object altering the propagation characteristics of radio frequency electromagnetic waves.} alter the propagation channel when they are close to the link-line of a TX-RX pair\footnote{We refer to the imaginary line connecting transmitter and receiver by \emph{link-line}.}, resulting in different RSS readings at the RX \cite{Patwari2010}. For such a localization system, there are four main components as shown in Fig.~\ref{fig:dfl-system}: \emph{models}, \emph{detector}, \emph{classifier} and \emph{location inference}. The complete system is built around the location inference method, which usually dictates the operation of the other components. The measurement pre-processing includes any signal conditioning and transformations required to form observations for the localization method. The models relate the observations to person's location, and they are altered by the classifier to improve the performance. A binary detector, whose operation also depends on the models, may be used to turn on and off the localization method. Therefore, a DFL task can be accomplished by a complex system, which requires development of different components.

In this work, our aim is to demonstrate that a lower-complexity localization system shown in Fig.~\ref{fig:system} can be used to realize a DFL system. The observations are compared against link-wise threshold values, which are determined using the model. The threshold values also define effective regions of each link, where they can identify presence of an interacting object. The classifier blacklists the links that cannot contribute to the localization effort. Binary output of the detector-bank is used by a back-projection based reconstruction algorithm that sums effective regions of the links indicating the presence of an interacting object.

The system shown in Fig.~\ref{fig:system} makes the following contributions compared to the traditional DFL system in Fig.~\ref{fig:dfl-system}. 
\begin{itemize}
\item The binary detectors require one to compare the current observation with a threshold value, and the reconstruction algorithm just sums the effective link regions. Thus, the computational complexity of the method is very low so that the complete localization method can be implemented in constrained microprocessors of wireless sensor nodes.
\item The binary detectors can be implemented on the receivers of each link since they do not require network-wise collaboration. Furthermore, the information needs to be transmitted is compressed to a single bit for each neighboring node. For the traditional system, the RSS measurements cannot be compressed, and a complete RSS value must be transmitted for each link limiting the scalability possibilities without complex network management. Hence, the proposed system enables highly scalable network deployments. 
\item We model the RSS variation using single-bounce reflections, which relates the threshold values to detection areas of each link. It has two parameters: \emph{path loss exponent} and \emph{reflection coefficient}. These two parameters have physical basis, and one can deduce their approximate values by making educated guesses. We show in the experimental evaluation section that these parameters are consistent among four different experiments in different environments, and yield stable detection thresholds. 
\end{itemize}

The remainder of the paper is organized as follows. In Section~\ref{sec:related-work}, the related literature is reviewed. The background material is given in Section~\ref{sec:background}, and in Section~\ref{sec:method} the methodology is introduced. The empirical evaluation of the system is given in Section~\ref{sec:results}. The conclusions are drawn in Section~\ref{sec:conclusion}.

\section{Related Work}\label{sec:related-work}
The reader is referred to the work by Patwari and Wilson \cite{Patwari2010} for a comprehensive overview  of the traditional localization system shown in Fig.~\ref{fig:dfl-system} until 2010.  Since our aim is to demonstrate the effectiveness of system in Fig.~\ref{fig:system}, we give a short literature review for each component. It is to be noted that not all works implicitly state the system components shown in Fig.~\ref{fig:dfl-system}. If a work does not utilize a classifier, the model is constant for the system under investigation. Similarly, if the system does not contain a detector, it implies either the method itself does not require explicit detection of person (e.g. radio tomographic imaging) or the person is always assumed to be in the region of interest. In this work, we do not consider the scenarios with more than one person is in the area of interest, and do not cover most of the important references on that subject. The reader is referred to the work by Wang et al. \cite{WangQ2015} and references therein.

The person's location can be inferred using a \emph{radio-map} \cite{Youssef2007, Moussa2009}, \emph{radio tomographic imaging} (RTI) \cite{Wilson2010}, or \emph{sequential Monte Carlo} (SMC) methods \cite{Li2011}. In a radio-map approach, a database is formed by collecting RSS values while person is standing at a location, for each location of the area, and in run-time the acquired RSS values are compared with the ones in that database. This approach requires significant effort in the preparation phase, and thus it is not considered within the scope of this paper. The SMC methods allow one to postulate the localization problem as a statistical inverse problem. They do not require discretization of the area of interest into pixels, and enable inclusion of movement dynamics into the formulation \cite{Li2011, Wilson2011a, Wilson2012}. Performance of the SMC method is mainly dictated by the number of samples drawn from posterior distribution in the Monte Carlo evaluations and the system modeling accuracy, including models of person's electrical properties, person's geometry and link-line geometry. In RTI, the RSS changes are used for estimating a map of the environment compared to static conditions \cite{Wilson2010}. RTI can operate even if the number of people in the environment is not known, or it is not possible or desirable to assume a detailed model for person's electrical properties and person's geometry. In RTI methods appeared so far, the map is reconstructed using an algebraic method \cite[ch. 7]{Kak1988}, which is a solution of an ill-posed linear inversion problem, demanding significant computational resources in creating a system model, and a matrix-vector product in run-time. In this regard, RTI has significantly lower computational requirements compared to SMC; yet, they can not be implemented on constrained microcontrollers, which do not usually possess a dedicated hardware for floating point arithmetic. Our focus in this paper is on developing a back-projection algorithm based RTI method, which enables such an implementation.


In the RTI literature, the RSS measurements are processed to obtain a linear observation model. The first model appeared assumes a large RSS decrease when the person is on the link-line \cite{Patwari2008}. Later, Wilson and Patwari have shown that the variation of RSS variance is linear with movements of a person \cite{Wilson2011a}, and they have pointed out that reflection and scattering are two mechanisms which can describe this variation \cite{Patwari2011}. For SMC based systems, on the other hand, more accurate models are needed. The empirical \emph{exponential} model accounts for the decay in logarithmic RSS measurements when the person is within the first Fresnel zone \cite{Li2011}. The \emph{Exponential Rayleigh} model includes first constructive peak of the RSS variation \cite{Guo2015}, which is a slight generalization of the exponential model. The knife-edge diffraction based models \cite{Rampa2015, Savazzi2014} require one to include (cylindrical) human model into consideration and assume that the human body is an absorber (with null reflection coefficient) in the first Fresnel zone. The work by Savazzi et al. \cite{Savazzi2014} assumes the person's centroid is on the link-line, and this restriction is relaxed in the work by Rampa et al. \cite{Rampa2015} to interior of the first Fresnel zone. The models alluded thus far are valid in a close neighborhood of the link-line, and they do not allow one to calculate  expected RSS when the person is outside of the considered region. The detector-bank in Fig.~\ref{fig:system}, however, requires a model which can relate the effective link regions to RSS values regardless of being in the first Fresnel zone. Although reflection based RSS model fulfills this requirement, it is valid if reflection is the dominating propagation mechanism. In this work, similar to the results of the work by Ghaddar et al. \cite{Ghaddar2007}, we demonstrate that reflection is the dominating mechanism, and use a reflection based model to build the system in Fig.~\ref{fig:system}.

Different DFL system components require calibration data in order to determine the operating parameters for an environment. This data can be acquired once and kept constant, or they can be modified in run-time of the system. For example, fade levels of the network links can be calculated using offline calibration data, whereas online methods, such as the ones studied in \cite{Kaltiokallio2012a} and \cite{Edelstein2013}, may alter them. In either case, they must be acquired when the system is in a specific state (e.g. when the monitored region is vacant). The proposed method requires constant calibration data to estimate the LoS signal power, and to blacklist the links in deep-fading state.

The DFL system can be incorporated with a classifier to improve their localization performance. Wilson and Patwari developed a classifier based on \emph{fade levels} of the network links for SMC based localization \cite{Wilson2012}. The fade level based classifier later enhanced by Kaltiokallio et al. \cite{Kaltiokallio2013} to improve the performance of RTI. The system shown in Fig.~\ref{fig:system} uses a classifier in order to identify and ignore the links that are in so-called \emph{deep-fade} state; i.e. to blacklist the links that cannot be used by the reconstruction algorithm.

Detecting presence of a person in the link's effective area is one of the outputs of DFL systems that can be utilized for occupancy assessment \cite{Naghiyev2014}. The systems proposed by Kosba \emph{et al.} \cite{Kosba2012} and Mrazovac \emph{et al.} \cite{Mrazovac2013} use a sequence of RSS measurements to make a decision. A recent work by Hillyard et al. detects line crossings using different classification techniques \cite{Hillyard2015}. In a work by Kaltiokallio and Bocca \cite{Kaltiokallio2011} link-line crossings are detected based on thresholds calculated in run-time, and occupancy is assigned to the pixels within an elliptic region. Similarly, Zheng and Men \cite{Zheng2012} have studied Gaussian mixture model based foreground detection. Different than the mentioned works, the proposed system is built around the link-level detectors based on the single-bounce reflection model, which relates effective regions of the links to the observation values. 

Energy efficient RSS based DFL system can be enabled by distributed processing \cite{Kaltiokallio2011}. Wu et al. \cite{Wu2016} has proposed a distributed RSS processing method, which allows representing a measurement with a single bit. It is argued that, in addition to distributed RSS processing, measurement compression is very important for energy efficient DFL system development. In this work, the distributed processing possibility is enabled by simplicity of the detectors, which do not require network-wise information but only observation associated with corresponding link. The output of each detector is a compressed representation of the link's measurement assigned by just comparing an observation to the threshold. Therefore, the proposed method enables low-complexity distributed RSS processing and measurement compression without introducing computation intensive operations. 

\begin{figure}[!t]
\centering
\setlength{\tabcolsep}{0.1cm}
\begin{tabular}{C{4.2cm}C{4.2cm}}
\subfloat[]{\includegraphics[width=4.1cm]{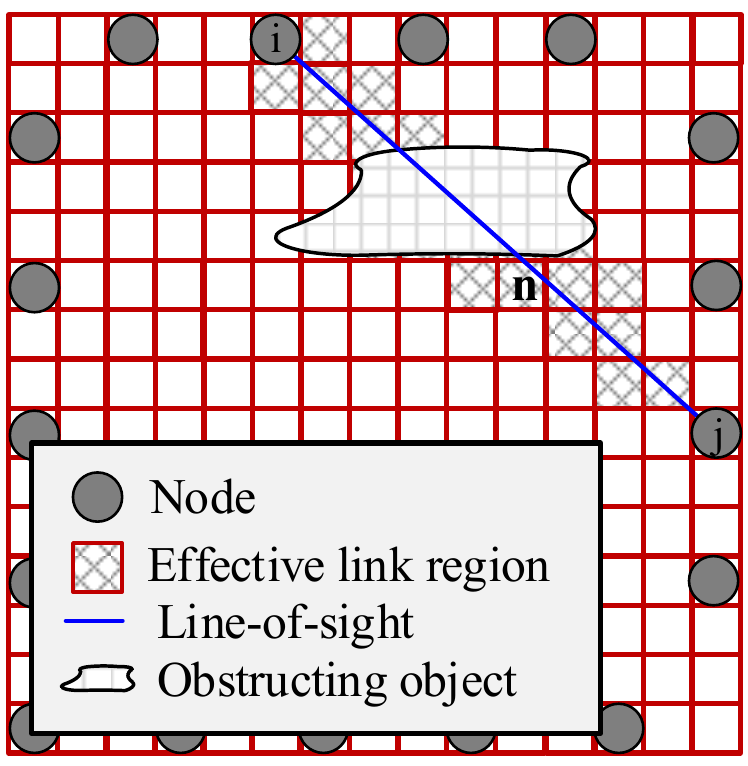}\label{fig:grid}} &
\subfloat[]{\includegraphics[width=4.1cm]{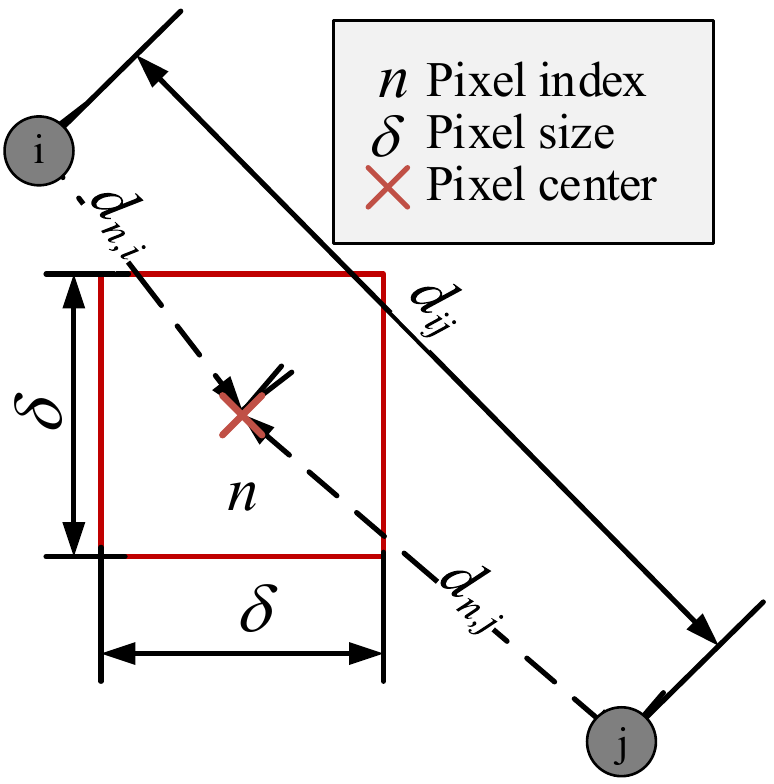}\label{fig:pixel}}
\end{tabular}
\caption{In (a), a typical Radio Tomographic Imaging grid and deployment scene. In (b), pixel level definitions. 
}
\label{fig:general_definitions}
\end{figure}

\section{Background} \label{sec:background}
In this section, necessary background material is given. We first define the general terms used throughout the paper, and state the relevant assumptions. Then, we introduce the measurement model. 

\subsection{General Definitions}\label{section:definitions}

We suppose that a low-power dense wireless network of $M$ nodes is deployed in a region of space as shown in Fig.~\ref{fig:general_definitions}. The nodes are programmed to communicate over $C$ communication channels, and the network is composed of $L$ links. We represent a link with a 3-tuple $(i,j,c)$ where $i$ the source identifier, $j$ is the destination identifier and $c$ is the channel identifier, but do not discriminate between links formed by the same nodes at different channels. The node identifiers are used to assign an unique identifier to each link in the network.

The region is discretized with a grid of $N$ pixels with equal lengths of $\delta \text{ m}$ in both dimensions as shown in Fig.~\ref{fig:general_definitions}. Let us denote the position of $i$\textsuperscript{th} node by $\boldsymbol{p}_i$ in an inertial reference frame in two dimensional Euclidean space. The center position of $n$\textsuperscript{th} pixel is denoted by $\boldsymbol{p}(n)$. The distance between $n$\textsuperscript{th} and $m$\textsuperscript{th} pixels is defined as $d(n, m) \triangleq \|\boldsymbol{p}(n) - \boldsymbol{p}(m)\|$, where $\|\cdot\|$ denotes Euclidean norm. Length of the line segment joining $i$\textsuperscript{th} and $j$\textsuperscript{th} nodes is denoted by $d_{ij}$. The distances between center of $n$\textsuperscript{th} pixel to the $i$\textsuperscript{th} and $j$\textsuperscript{th} nodes are denoted by $d_{n,i}$ and $d_{n,j}$ in respective order as visualized in Fig.~\ref{fig:pixel}. The excess path length traversed by a ray if it would be reflected from the center of $n$\textsuperscript{th} pixel is given by
$
	\Delta_{n, ij} \triangleq d_{n,i} + d_{n,j} - d_{ij}.
$
The notation is simplified when only one transmit-receiver pair is considered. In particular, subscript $_{ij}$ is dropped from $d_{ij} \equiv d = \|\boldsymbol{p}_r - \boldsymbol{p}_t\|$, where $\boldsymbol{p}_t$ denotes position of transmitter and $\boldsymbol{p}_r$ denotes position of the receiver.

For the studied reflection model, ellipses play an important role. Suppose that an ellipse has its foci at $\boldsymbol{p}_r$ and $\boldsymbol{p}_t$, which are separated by $d$. In this case, any point $\boldsymbol{p}$ on the ellipse can be parametrized using a single parameter $\Delta$, 
\begin{equation}\label{eq:excess_path_length}
	\Delta \triangleq \|\boldsymbol{p} - \boldsymbol{p}_r\| + \| \boldsymbol{p} - \boldsymbol{p}_t\| - d.
\end{equation}
In terms of $\Delta$ and $d$, the area of an ellipse is given by
\begin{equation}\label{eq:excess_path_length_ellipse_area}
	A(d, \Delta) \triangleq \frac{\pi}{4}(d + \Delta)\sqrt{2d\Delta + \Delta^2}.
\end{equation}

\begin{figure}[t!]
\centering
\subfloat[]{\includegraphics[width=6cm]{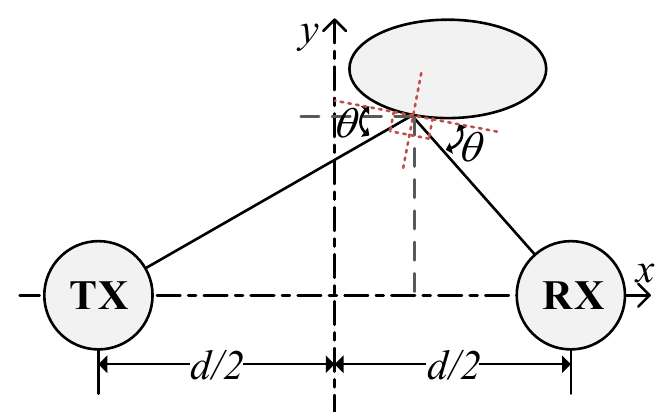}}
\caption{The propagation scenario for single-bounce reflection}
\label{fig:reflection_geometry}
\end{figure}

\subsection{RSS Measurement Model}\label{sec:measurement-model}
Consider the radio wave propagation scenario depicted in Fig.~\ref{fig:reflection_geometry}, where the propagation environment is completely empty except the interacting object and the TX-RX pair. Suppose that the communication system satisfies the following: 
\begin{inparaenum}[i.)] 
\item{The wavelength is much smaller than average geometrical extent of the interacting object.}
\item{The antennas have ideal frequency response and gain patterns, and they are vertically polarized.}
\end{inparaenum}

Under these assumptions, some of the electromagnetic energy emitted by the TX is reflected from the surface of the dielectric interacting object and, there are two multipath components impinging on the receiver antenna: the LoS signal and the single-bounce reflected signal. The received complex baseband signal in such a scenario can be written as \cite[ch.~13]{Proakis2008} 
\begin{equation*}
	r(t) = \sum\limits_{i=1}^{2}{\alpha_i(t)e^{-j 2\pi f_c \tau_i}s(t-\tau_i)} + \hat{n}(t), 
\end{equation*}
where $f_c$ is the center frequency, $\alpha_i(t)$ and $\tau_i$ are amplitude and time delay of $i$\textsuperscript{th} multipath component in respective order, $\hat{n}(t)$ is complex zero-mean noise with two-sided power spectral density $N_0/2$, and $s(t)$ is the transmitted, which has symbol duration $T_s$. If the LoS signal arrives to the receiver antenna first, $\tau_1$ is zero and for the systems with no inter symbol interference we have $0 < \tau_2 \ll T_s$. If we further suppose that the propagation medium is constant for the time period of interest, the amplitudes of individual components are constant, and the received signal is given by
\begin{equation}\label{eq:received_signal}
	r(t) = \alpha_1s(t) + \alpha_2 e^{-j \phi}s(t-\tau_2) + \hat{n}(t), 
\end{equation}
where $\phi = 2 \pi f_c \tau_2 = 2 \pi \Delta f_c / c_0$ and $c_0$ is the free-space electromagnetic propagation speed. For a quadrature modulation scheme with the same in-phase and quadrature shaping pulses and when $\hat{n}(t)$ is mean ergodic, the received signal power calculated for a large time window $T \gg T_s$ and converted to logarithmic scale is given by
\begin{equation}\label{eq:received_power}
	\begin{aligned}
	\mathcal{P}_r = \mathcal{P}_s + & 10 \log_{10}(\alpha_1^2) + 10 \log_{10}\left(1 + {\alpha_2^2}/{\alpha_1^2}\right) + \\ 
					& 10 \log_{10}\left(1+\frac{2 \alpha_1 \alpha_2}{\alpha_1^2 + \alpha_2^2 } \cos(2 \pi \Delta / \lambda)\right) + {n},
	\end{aligned}
\end{equation}
where $\lambda \triangleq f_c / c_0$ is the wavelength of the emitted electromagnetic waves, $\mathcal{P}_s$ is the transmitted signal power and ${n}$ denotes power measurement noise in logarithmic scale.

The ratio between electric field intensities of incident and reflected waves is given by the Fresnel reflection coefficient \cite[ch. 4]{Rappaport2002}. The coefficient depends on polarization of the incident wave with respect to the plane of incidence, incidence angle $\theta$, electrical properties of the reflection surface, and the wavelength. It is a real number when the object is dielectric or a complex number if the object has non-zero conductivity. The absolute value of the reflection coefficient is defined by electrical properties of the dielectric boundary. Considering humans, clothing is the dielectric boundary, and (dry) human skin is the second reflection layer. Therefore, an accurate model for reflection can be developed by taking into account antenna polarization and gain pattern, electrical properties of multiples of dielectric boundaries, human body geometry and position with respect to link-line.

The relative permittivity of common textile materials are $\epsilon_r = 1.5 - 2$ in $2.4 \text{ -- } 2.5 \text{ GHz}$ \cite{Sankaralingam2010}, and for human skin $\epsilon_r = 38$ in the same frequency band \cite{TissueElectric}. Such values of $\epsilon_r$ implies high reflected RF power for all incidence angles. Furthermore, when the TX antenna is standing in the direction perpendicular to plane of incidence ($z$ axis in Fig.~\ref{fig:reflection_geometry}), the electric field intensity of the emitted wave is oriented along the same axis so that the reflected waves only have perpendicular polarization, which yields negative Fresnel reflection coefficient for all incidence angles. In what follows, we assume that the ratio between reflected signal amplitude and incident signal amplitude $\Gamma$ is a non-negative real scalar satisfying $0.1 \le \Gamma < 0.75 $ and suppose that sign of the Fresnel reflection coefficient contributes $\pi$ radians to phase of the reflected signal. We further do not consider the position and orientation dependence of Fresnel reflection coefficient. These assumptions imply that $\Gamma$ is zero order approximation of the (perpendicular) Fresnel reflection coefficient, and throughout remainder of the paper, we refer to it as \emph{reflection coefficient}.

The amplitude of both multipath components in Eq.~\eqref{eq:received_power} have the same decay rate with distance if both experience the same realization of fading, which can be modeled as a cite-dependent path-loss exponent $\eta$. In this case, the amplitudes of the components decrease with $(-\eta/2)^{\text{th}}$ exponent of the traversed path-length. Combining this with the reflection coefficient $\Gamma$ defined in the previous paragraph yields
\begin{equation*} \label{eq:amplitude_assumptions} 
		\alpha_1 \approx  d^{-\eta/2}, \qquad
		\alpha_2 \approx -\Gamma \left(d+\Delta \right)^{-\eta/2},
\end{equation*}
where $\Delta$ is the excess path length traversed by the reflected signal. Therefore, the received signal power in Eq.~\eqref{eq:received_power} can be written as
\begin{equation*} \label{eq:received_power2}
	\mathcal{P}_r = \mathcal{P}_0 + \zeta(\Delta, \Gamma, d, \eta, f_c) + {n},
\end{equation*}
where $\mathcal{P}_0$ is the LoS signal power and the effect of the reflected multipath on the RSS measurement is defined as
\begin{equation}\label{eq:effect_of_reflection_db}
\begin{aligned}
	\zeta(\Delta, \Gamma, d, &\eta, f_c) \triangleq  10 \log_{10}\left(1+\frac{\Gamma^2}{(1+{\Delta}/{d})^{\eta}}\right) + \\
			& 10\log_{10}\left(1 - 2\Gamma\frac{(1+\Delta/d)^{\frac{\eta}{2}}}{\Gamma^2  + (1+\Delta/d)^\eta} \cos(\phi)\right).
\end{aligned}
\end{equation}
In sequel, we do not explicitly state $\Gamma$, $d$, $\eta$ and $f_c$ dependence of $\zeta(\cdot, \cdot, \cdot)$ function, and write it as $\zeta(\Delta) \equiv \zeta(\Delta, \Gamma, d, \eta, f_c)$. 

\begin{figure*}[!t]
\centering
\setlength{\tabcolsep}{0.1cm}
\begin{tabular}{C{5.7cm}C{5.7cm}C{6.2cm}}
\subfloat[]{\includegraphics[width=5.6cm]{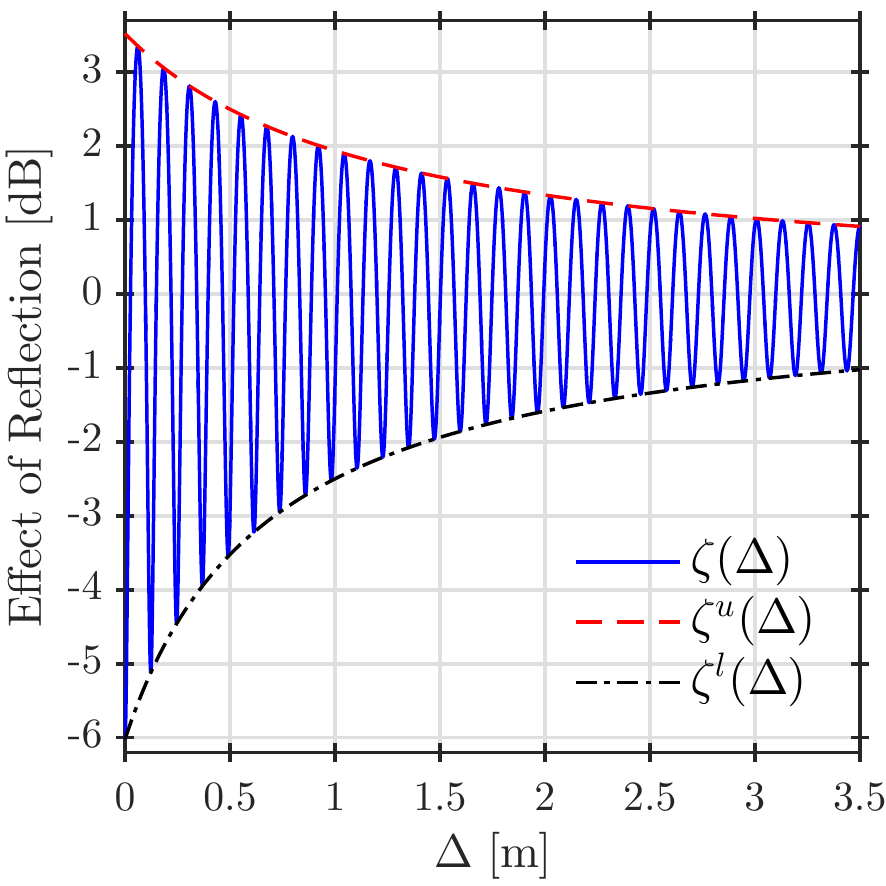}\label{fig:reflection}} &
\subfloat[]{\includegraphics[width=5.6cm]{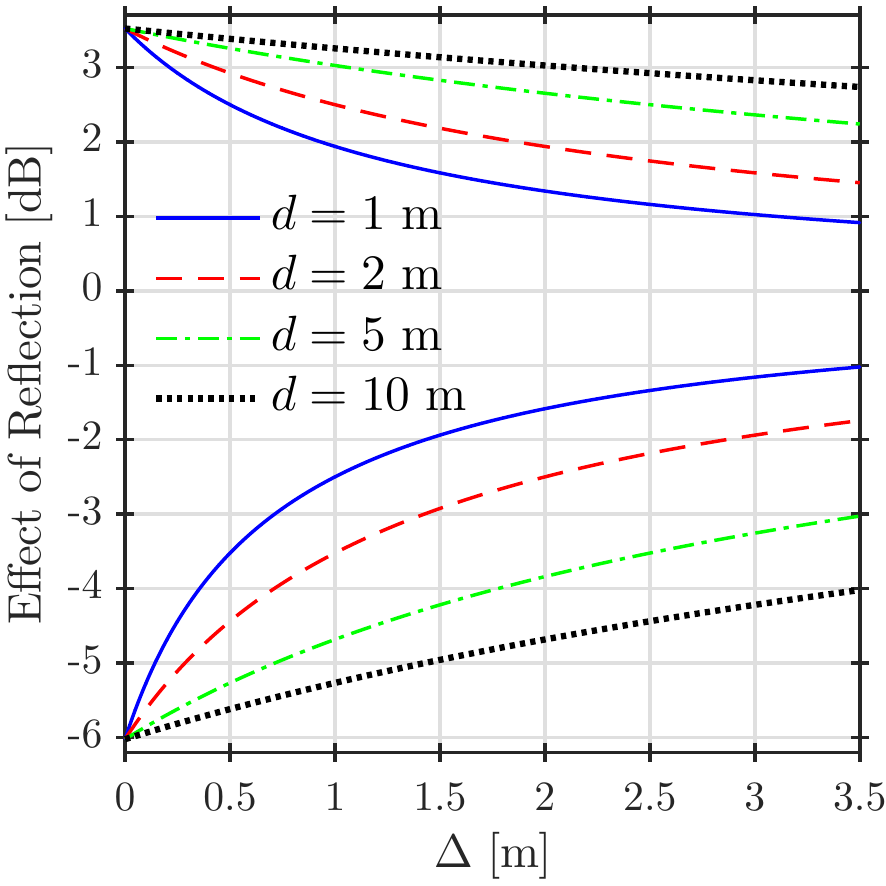}\label{fig:reflection-d}}&
\subfloat[]{\includegraphics[width=6.2cm]{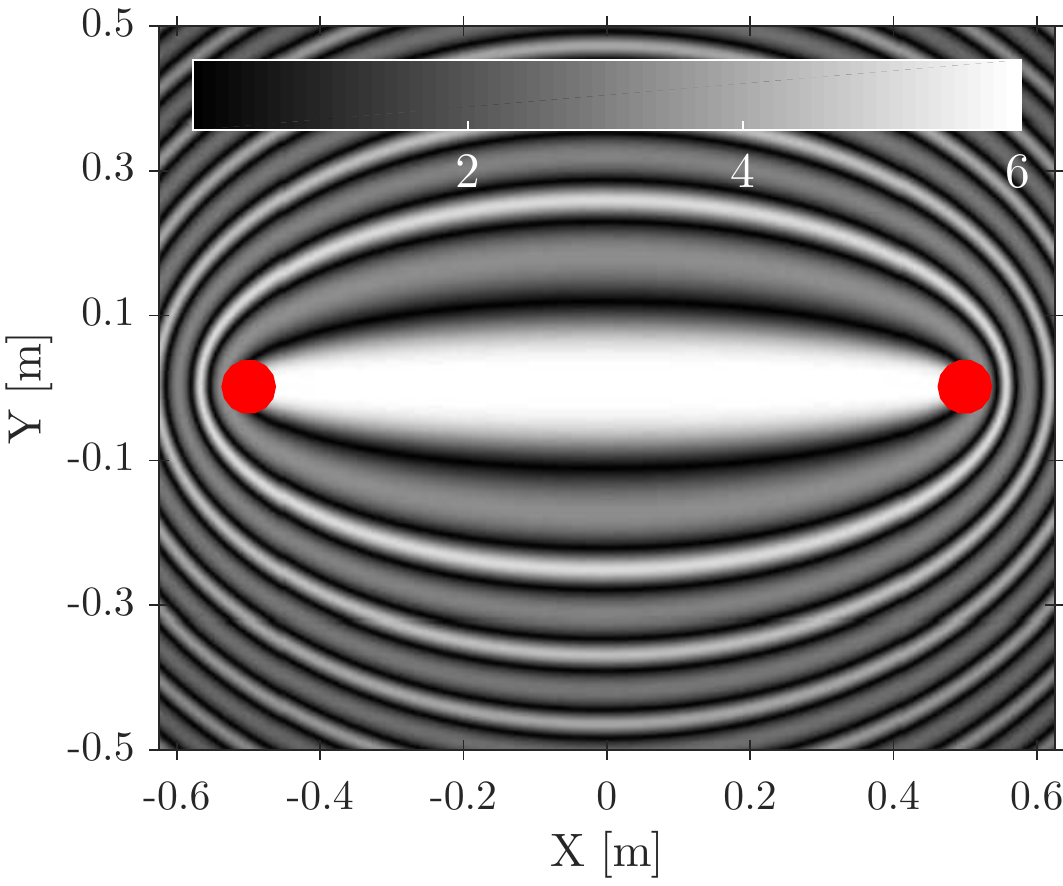}\label{fig:reflection-2d}}
\end{tabular}
\caption{In (a), the effect of single-bounce reflection on the RSS defined in Eq.~\eqref{eq:effect_of_reflection_db} and its envelopes in Eq.~\eqref{eq:zeta-envelopes} for $\Gamma = 0.5$, $\eta = 2.0$, $d=1.0 \text{ m}$ at $2.4425 \text{ GHz}$. In (b), the variations of envelope functions for different $d$ values. In (c), variation of $|\zeta(\Delta)|$ with position for $\Gamma = 0.5$, $\eta = 2.0$ at $2.4425 \text{ GHz}$ and nodes at $[-0.5, \,0]^T$ and $[0.5, \,0]^T$ (shown as red circles).
}
\label{fig:reflection_model}
\end{figure*}

The derivations for the ideal scenario can be extended to cover more general conditions by including additional losses encountered in practical deployments. In logarithmic scale, these effects can be represented by a frequency dependent parameter $\mathcal{P}_0(f_c)$ and additional stochastic parameters can be added to measurement noise. Hence, the RSS measurement in logarithmic-scale can be written as 
\begin{equation*}\label{eq:received_power_db1}
\mathcal{P}_r(\Delta, f_c) = \mathcal{P}_{0}(f_c) + \zeta(\Delta) + \hat{\nu}(f_c), 
\end{equation*} 
where $\hat{\nu}(f_c)$ is the measurement noise. Since $\mathcal{P}_{0}(f_c) + \mathrm{E}\{\hat{\nu}(f_c)\}$ is the most probable received signal power measurement (when $\hat{\nu}(f_c)$ has a unimodal density), it can be estimated before the person enters to the monitored area, and then subtracted from all the measurements. Consequently, an observation model for $\mathcal{P}_r(\Delta, f_c) - \mathcal{P}_{0}(f_c) - \mathrm{E}\{\hat{\nu}(f_c)\}$ is given by
\begin{equation}\label{eq:observation-model}
	{z} \triangleq \zeta(\Delta) + \nu(f_c).
\end{equation}

\section{Method} \label{sec:method}
In this section, components of the system shown in Fig.~\ref{fig:system} are introduced. We elaborate the implications of the measurement model derived in the previous section. The link-level detectors and the link black-listing classifier are introduced after the model. Next, the back-projection based reconstruction algorithm is presented, which uses the output of these two subsystems. The location estimator is given before we discuss the practicalities associated with the methods introduced in this section.

\subsection{Single-bounce Reflection}\label{section:reflection-model}
The effect of reflection function $\zeta(\Delta)$ in Eq.~\eqref{eq:effect_of_reflection_db} has different upper ($\zeta^u(\Delta)$) and lower ($\zeta^l(\Delta)$) envelopes,
\begin{subequations}\label{eq:zeta-envelopes}
\begin{align}
\zeta^u(\Delta) &= 20\log_{10} \left(1 + \frac{\Gamma}{(1+\Delta/d)^\frac{\eta}{2}}\right)\label{eq:zeta_envelope_upper}\\
\zeta^l(\Delta) &= 20\log_{10} \left(1 - \frac{\Gamma}{(1+\Delta/d)^\frac{\eta}{2}}\right)\label{eq:zeta_envelope_lower}
\end{align}
\end{subequations}
which are $\zeta(\Delta) = \zeta^u(\Delta)$ when $\cos(\phi)=-1$ and $\zeta(\Delta) = \zeta^l(\Delta)$ when $\cos(\phi)= 1$. If $\Delta \ll d$, the first two term of Maclauren series expansion of the logarithmic envelope functions in Eq.~\eqref{eq:zeta-envelopes} can be written as
\begin{equation*}
\begin{aligned}
\zeta^u(\Delta) &\approx \frac{20}{\ln({10})} \left(\ln(1+\Gamma) - \frac{\eta}{2}\frac{\Gamma}{\Gamma+1}\frac{\Delta}{d}\right), \\
\zeta^l(\Delta) &\approx \frac{20}{\ln({10})}\left(\ln(1-\Gamma) + \frac{\eta}{2}\frac{\Gamma}{\Gamma+1}\frac{\Delta}{d}\right),
\end{aligned}
\end{equation*}
which show that the rate of change of envelope functions with $\eta$ are very small. Thus, for fixed TX and RX locations, $\Gamma$ is the primary parameter which defines the maximum variation of the RSS due to reflections and one should investigate the variation of $\zeta(\Delta)$ with $\Delta$. 

\subsubsection{Numerical Evaluation}
In Fig.~\ref{fig:reflection}, the variation of $\zeta(\Delta)$ with $\Delta$ for $\Gamma=0.25$, $\eta=2.0$ and $d = 1.0 \text{ m}$ at $f_c=2.4425 \text{ GHz}$ is plotted along with its upper and lower envelopes. For the same parameters, the variation of envelope functions with $\Delta$ and different $d$ values are plotted in Fig.~\ref{fig:reflection-d}. The figure visualizes almost linear variation of envelope functions with $\Delta$ for large $d$ values. The absolute value of effect of reflection on RSS measurement $|\zeta(\Delta)|$ has larger regions above certain threshold when $\Delta$ is small as shown in Fig.~\ref{fig:reflection-2d}. Therefore, when $\Delta \ll d$, the envelopes of $\zeta(\Delta)$ are almost linear so that for given parameters ($\Gamma$, $d$, $\eta$) the envelope values can be used to assess expected extreme values of the observations for a given excess path length $\Delta$.  

\begin{figure*}[!t]
\centering
\setlength{\tabcolsep}{0.1cm}
\begin{tabular}{C{3cm}C{7.25cm}C{7.25cm}}
\subfloat[]{\includegraphics[width=2.8cm, height=6cm]{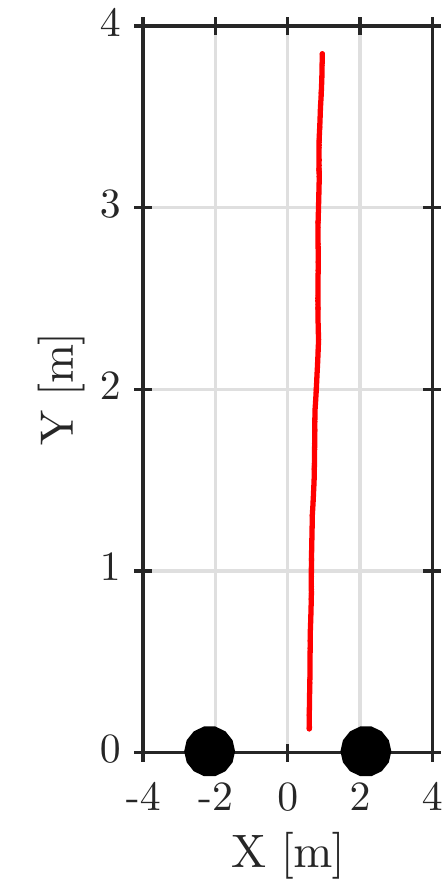}\label{fig:experiment0}} &
\subfloat[]{\includegraphics[width=7.2cm, height=6cm]{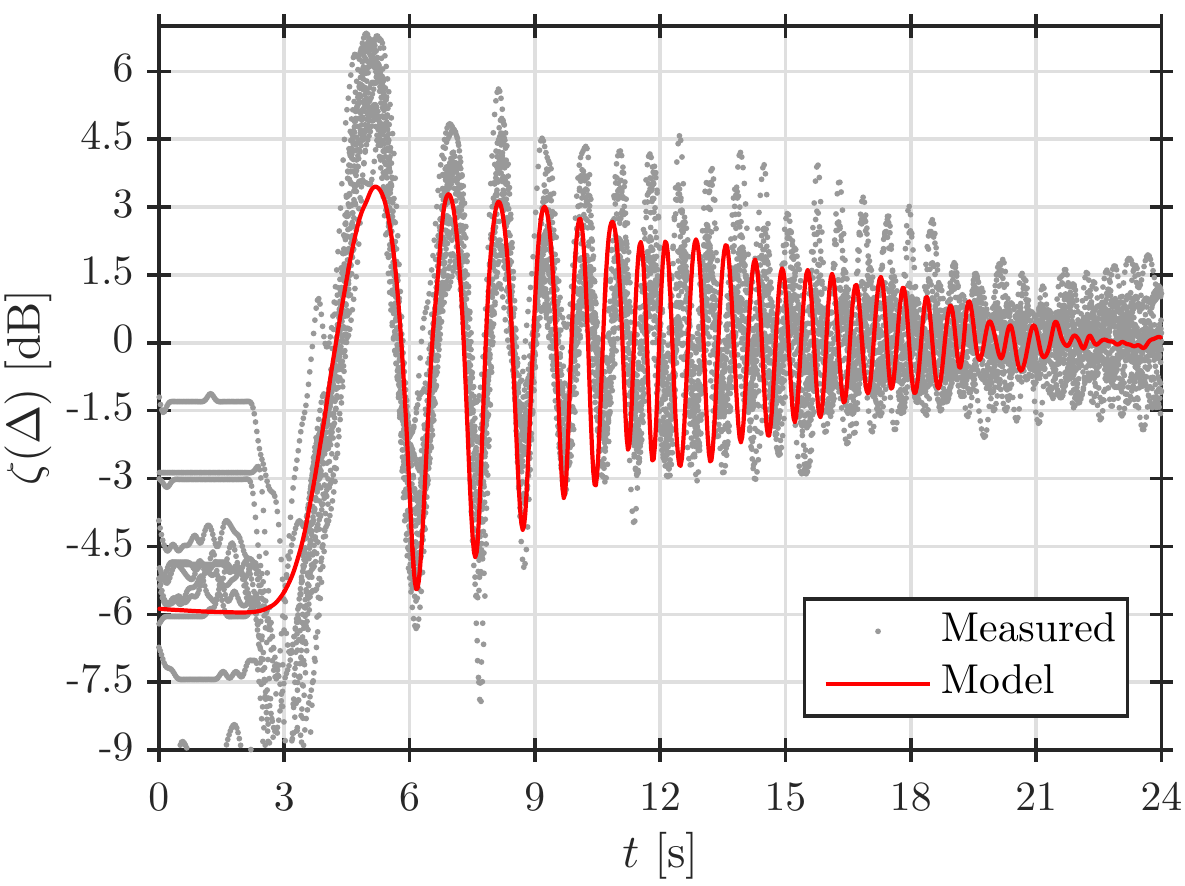}\label{fig:experiment-time}}&
\subfloat[]{\includegraphics[width=7.2cm, height=6cm]{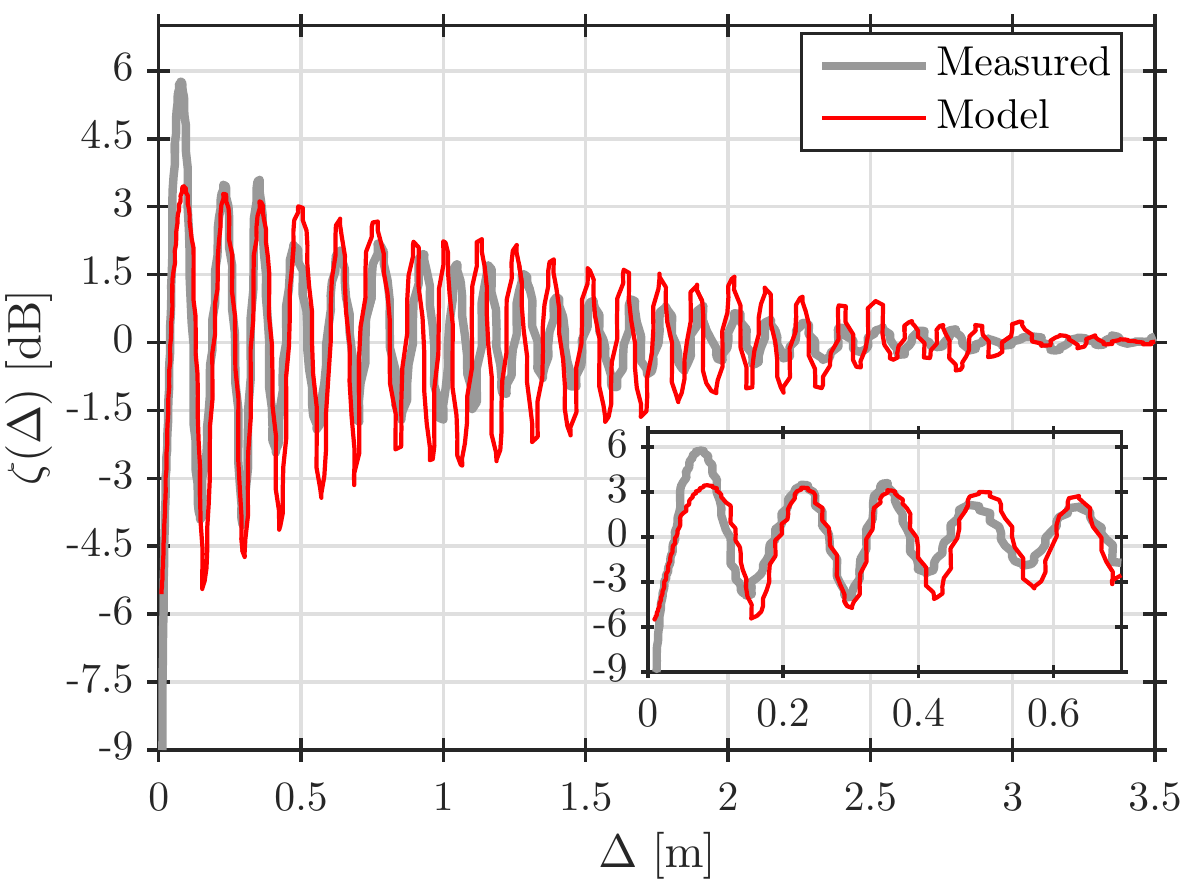}\label{fig:experiment-delta}}
\end{tabular}
\caption{In (a), the placement of the nodes (shown as black dots) and robot movement trajectory (the red line) towards increasing $Y$ coordinate. In (b), a comparison between acquired RSS time samples in different channels with the model calculated values averaged over all channels for $\Gamma = 0.5$, $\eta = 2.0$ and using robot pose estimates. In (c), corresponding result when the robot position (using known container geometry) is converted to excess path length $\Delta$ and averaged over all channels.
}
\label{fig:reflection-model-experiment}
\end{figure*}

\subsubsection{Experimental Validation}
In order to validate the reflection model, a robot equipped with a cylindrical container filled with special liquid imitating electromagnetic properties of the human body, explained in Sec.\ref{section:result-experiments}, is used. Only a single transmitter-receiver pair is deployed with the same hardware components used in Experiment~\Rom{1} in Sec.\ref{section:result-experiments} except both transmitter and receiver are equipped with directional antennas with $75^\circ$ horizontal beam width.

In the experiment, the transmitter and receiver are placed $ 4.282 \text{ m}$ apart from each other, and the antennas are rotated in order to allow reception of reflected signal from the robot at all positions in the region of interest. The transmitter emits frames every $1.92 \text{ ms}$ at all $16$ channels of IEEE $802.15.4$ MAC/PHY specification at $2.4 \text{ GHZ}$ band in Round-Robin fashion. Each transmitted frame contains a unique frame identifier so that the receiver can identify lost frames and the robot pose logs can be synchronized to transmissions. The robot is navigated remotely on a linear trajectory by commanding it over a $5 \text{ GHz}$ wireless link. The deployment scene and trajectory of the robot are shown in Fig.~\ref{fig:experiment0}. 

The RSS values acquired by the RX node are subtracted from estimated LoS signal power, and then filtered with a maximally flat finite-impulse-response filter with $3 \text{ dB}$ cut-off frequency of $8 \text{ Hz}$. The reflection point on the container is calculated solving the non-linear kinematic equations in two dimensions using the inertial measurements and pose estimates of the robot. Then, the reflection points are used for calculating the excess path length. The above mentioned filter is also used on the model output for the calculated excess path length values. This filter improves visual quality of the results by decreasing effect of the quantization while preserving spectral properties of measurements since the robot speed is low ($ < 0.3 \text{ m/s}$).

Time series of the processed RSS at all channels are shown in Fig.~\ref{fig:experiment-time}. The model output $\zeta(\Delta)$ for excess path length $\Delta$ values of the container are calculated for each channel and then averaged over all the channels is shown on the same figure. The averaging is used for simplifying the plot since different channels have significantly different phase when excess path length is large. The same result is also shown in Fig.~\ref{fig:experiment-delta} as a function of excess path length $\Delta$ and when the processed RSS values are also averaged over all the channels. These results are coherent with findings of Ghaddar et al. \cite{Ghaddar2007}, and one can conclude that single bounce reflection model closely resembles the RSS variations.

\subsection{Link-level Detector} \label{sec:detector}
For the observation model given in Eq.~\eqref{eq:observation-model}, presence of the reflected signal ($\mathcal{H}_1$) can be tested against absence of it ($\mathcal{H}_0$):
\begin{equation*}
	 \mathcal{H}_0: {z} = {\nu}, \qquad
	 \mathcal{H}_1: {z} =  \zeta(\Delta) + {\nu},
\end{equation*}
in order to identify presence of an interacting object. Such a detector is referred as binary hypothesis test \cite[ch. 2]{Trees1968}, and it makes a decision based on a threshold associated with the decision-criteria. Its performance for a given threshold can be evaluated using the statistical model of ${z}$, which is dictated by the noise term $\nu(f_c)$ in Eq.~\eqref{eq:observation-model}. If we suppose that noise $n$ in Eq.~\eqref{eq:received_power} is the dominating factor, we can approximate $\nu \approx n -\mu_{n,h}$ ($\mu_{n,h}$ is the mean of $n$ when hypothesis $\mathcal{H}_h$ is true). When the power is calculated in digital domain using $K$ samples of the received signal, the relevant statistical properties of $n$ for additive white Gaussian noise process $\hat{n}(t)$ are derived in Appendix~\ref{appendix:noise-derivations}. 

The number of samples used in power calculations $K$ is very large for direct sequence spread spectrum systems such as IEEE~802.15.4 standard compliant radios. For example, these radios spread their information symbols over $32$ chips, and power calculations are performed over $8$ symbol periods \cite{802_15_4}, which means $K \ge 512$ (multiplied by $2$ due to Nyquist sampling criteria). Such large values of $K$ implies that $n$ approaches to its asymptotic distribution derived in Appendix~\ref{appendix:noise-derivations}. As $K \to \infty$, $n$ approaches to the constant value in Eq.~\eqref{eq:n-asymptotic-mean}, but the observed variations are due to narrow dynamic range quantization of the power measurements. Therefore, if one restricts the effective regions of links by stating maximum excess path length as $\Delta_t$ that yields absolute envelope values in Eq.~\eqref{eq:zeta-envelopes} much larger than the quantization levels, she can select the detection thresholds using the envelope values. In this case $\mathcal{Z} = |\zeta^{l}(\Delta_t)| $, which is the maximum of $|\zeta^{u}(\Delta_t)|$ and $|\zeta^{l}(\Delta_t)| $, can be used as a threshold value. Consequently, in this work, we define detection threshold for a link and make the decisions by comparing the observations with it, i.e.
\begin{equation}\label{eq:decision}
	|z| \underset{\mathcal{H}_0}{\overset{\mathcal{H}_1}{\gtrless}} \mathcal{Z}, \qquad \mathcal{Z} = |\zeta^{l}(\Delta_t)|.
\end{equation}

\subsection{Link Classifier}

Wireless links may exhibit non-ideal behavior depending on different factors such as radiation patterns of the antennas, frequency response of the analog amplifiers and filters. These factors significantly lower the received signal power compared to losses predicated by log-distance path loss propagation model (see e.g. \cite[ch. 4]{Rappaport2002}). In the literature, the amount of deviation of RSS measurements from the model predicted value is defined as \emph{fade-level} of a link \cite{Wilson2012, Kaltiokallio2013} and significantly low fade-levels are referred as \emph{deep-fading} \cite{Wilson2012}. 

The log-distance propagation model is defined as 
\begin{equation}\label{eq:log-distance-model}
	\mathcal{P}_r = \mathcal{P}_s - \mathcal{P}_1 - 10 \eta \log_{10}\left( {d}/{d_1} \right) + n,
\end{equation}
where $\mathcal{P}_1$ is the average signal power measurement at a reference distance $d_1$. By comparing the received signal power in Eq.~\eqref{eq:received_power} when the interacting object is not in the environment ($\zeta(\Delta) = 0$) with the log-distance model in Eq.~\eqref{eq:log-distance-model}, one can deduce that
\begin{equation}
	\mathcal{P}_0 = \mathcal{P}_s - \mathcal{P}_1 - 10 \eta \log_{10}\left( {d}/{d_1} \right).
\label{eq:P0-log-distance}
\end{equation}
The average power measurement at the reference distance $d_1$ and the transmission power are constant, and $\eta$ is the path-loss exponent of the received signal power averaged over all link-line directions so that it is fixed for a site. Therefore, a link may be in deep-fade only if there is an uncertainty in any of the parameters. 


The RSS measurements of deeply-faded links can not be used by the detector to identify presence of the interacting object since distance traversed by the reference signal (signal with zero phase in Eq,~\eqref{eq:received_signal}) and associated access path lengths are not deterministic quantities. This problem can be avoided by black-listing the deeply faded links. For this purpose, the link level classifier compares the RSS measurement when the interacting object is not in the environment (i.e. $\mathcal{P}_r = \mathcal{P}_0 + \nu(f_c)$) with the log-distance model predicted value and outputs a decision. Thus, a link is black-listed as follows
\begin{equation}
	\iota = 
	\begin{cases}
		0 & \mathcal{P}_r - \left(\mathcal{P}_s - \mathcal{P}_1 - 10 \eta \log_{10}\left( {d}/{d_1} \right) \right) \le \mathcal{P}_d  \\
		1 & \text{otherwise}
	\end{cases}
\label{eq:black-listing}
\end{equation}
where $\mathcal{P}_d$ is a negative threshold value which must be specified according to minimum dip of the RSS. 

\subsection{Back-projection Image Reconstruction Algorithm}
The tomographic imaging of non-diffracting sources using back-projection reconstruction is a weighted sum of the measurements assigned to the lines traversed by the rays \cite[ch. 2]{Kak1988}. Similar to the binary detector's output, these systems may use binary valued measurements (possibly indicating sum of attenuation values above a threshold), which can indicate presence of, e.g., bones on line traversed by a ray. Any similar reconstruction algorithm, therefore, requires two parameters: 
\begin{inparaenum}[i.)]
\item the effective region for a measurement (e.g. the line of the ray), and
\item the weight associated with each pixel.
\end{inparaenum}

The localization algorithm studied in this work is analogous to back-projection reconstruction, where considered effective regions for the measurements are ellipses instead of lines. Since ellipses are two dimensional, the effective regions must be restrained by defining a maximum excess path length value $\Delta_t$. For this case, an indicator matrix, which has non-zero entries for pixels effecting the link RSS measurements, can be defined as
\begin{equation}
\boldsymbol{\mathbbm{1}}_{n, ij} =
\begin{cases}
	1 & \Delta_{n,ij} \le \Delta_t \\
	0 & \text{otherwise}
\end{cases}
\label{eq:reflection-indicator}
\end{equation}
where subscript $_{n,ij}$ indicates a value at $n$\textsuperscript{th} pixel for a link between $i$\textsuperscript{th} transmitter and $j$\textsuperscript{th} receiver. The excess path length $\Delta$ parametrization of the ellipses allows us to use the same threshold value for different links while having different observation thresholds in Eq.~\eqref{eq:decision}.

The binary output of the detectors indicate whether there is a reflection point on the interacting object's exterior which yields observations greater than $\mathcal{Z}$ as discussed in Sec.~\ref{sec:detector}. In other words, if a link indicates presence, the object can be at any point (pixel) with equal probability in the link's effective region, yielding a weight model ${1}/{A(d_{ij}, \Delta_t)}$, where $A(\cdot, \cdot)$ is defined in Eq.~\eqref{eq:excess_path_length_ellipse_area}. Different links, on the other hand, may measure occupancy at the same positions making some pixels indicate occupancy multiple times. Each pixel may have equal contribution to indicate presence of an interacting object using the following scale,
\begin{equation}
	\mathbbm{S}_{n,ij} = 
	\begin{cases}
		\frac{1}{A(d_{ij}, \Delta_t)}\frac{1}{\sum_{ij}\{1/A(d_{ij}, \Delta_t)\}} &\Delta_{n,ij} \le \Delta_t \\
		0 & \text{otherwise}
	\end{cases}
\label{eq:reflection-scale}
\end{equation}
These scale values sum to one when all the links containing the $n$\textsuperscript{th} pixel in their effective regions indicate presence of the object. We refer to sum of these scales as \emph{occupancy of a pixel}. The reconstruction algorithm makes use of a weight matrix $\boldsymbol{W} \in \mathbbm{R}^{N\times L}$ defined as
\begin{equation}\label{eq:weight-matrix-decomposition}
	\boldsymbol{W} = \boldsymbol{\mathbbm{1}} \odot \boldsymbol{\mathbbm{S}},
\end{equation} 
where $\odot$ is element-wise (Hadamard) matrix multiplication. Although we have $\boldsymbol{W} = \boldsymbol{\mathbbm{S}}$ since $\mathbbm{S}_{n,ij} = 0$ when $\boldsymbol{\mathbbm{1}}_{n, ij} = 0$, we later show in Sec.~\ref{section:result-discussion} that in fact $\boldsymbol{\mathbbm{S}}$ can be calculated from binary values of $\boldsymbol{\mathbbm{1}}$.

Let $\boldsymbol{\xi} \in \mathbbm{R}^L$ denote the binary output vector of the detectors for each transmit receive pair (we later discuss the links at different channels in Sec.~\ref{section:method-discussion}). Similarly, let $\boldsymbol{\iota} \in \mathbbm{R}^L$ denote the output of the classifier for each transmitter and receiver pair with components defined in Eq.~\eqref{eq:black-listing}, and $\boldsymbol{\mathcal{I}} \triangleq \diag\{\boldsymbol{\iota}\}$, where each component of $\boldsymbol{\iota}$ is placed on the diagonal of $\boldsymbol{\mathcal{I}}$. Then, a spatial occupancy field is given by
\begin{equation}\label{eq:probability-field}
	\hat{\boldsymbol{\Pi}} = \boldsymbol{W} \boldsymbol{\mathcal{I}} \boldsymbol{\xi}.
\end{equation}
Hence, location of the interacting object can be determined by just summing weights of the links that are not in deep-fade and detecting object's presence. Suppose there are $L^\prime \le L$ such links at a time. Then, the algorithm requires $L^\prime$ floating point additions for each pixel, so that occupancy field of the monitored area is formed by $L^\prime N$ additions.

\subsection{Location Estimation}
The spatial field of interacting object presence given in Eq.~\eqref{eq:probability-field} estimates the occupancy of each pixel. If one knows that there is only one interacting object in the environment, it must be surrounded by the pixels with the highest occupancy, i.e., the regions around the mode of the field $\hat{\boldsymbol{\Pi}}$. 
Therefore, the location estimation can be postulated as finding the regions containing the modes of the spatial field $\hat{\boldsymbol{\Pi}}$.

Let $\mathbb{P}^*$ denote the maximum component of $\hat{\boldsymbol{\Pi}}$. Then, the mode is in the set of pixels indexes $\{ i: \hat{\Pi}_i \ge a\mathbb{P}^*\}$, which is the region composed of pixels with occupancy higher than $a\mathbb{P}^*$ for $0 < a < 1$, possibly covering exterior of the interacting object. 
Thus, a good location estimate is at the weighted sum of these pixels. For notational convenience let us define 
\begin{equation}\label{eq:probability-field2}
	\bar{\Pi}_n = 
	\begin{cases}
		\hat{\Pi}_n & \hat{\Pi}_n \ge a\mathbb{P}^*\} \\
		0 & \text{otherwise}
	\end{cases}
\end{equation}
and $\tilde{\boldsymbol{\Pi}} = \bar{\boldsymbol{\Pi}} / \left(\sum_n \bar{{\Pi}}_n \right)$. Then, a position estimate is given by
\begin{equation}
	\hat{\boldsymbol{p}} \triangleq \sum\limits_{n=1}^{N}\boldsymbol{p}(n)\tilde{\Pi}_n.
\label{eq:position-estimate}
\end{equation}

\subsection{Discussion}\label{section:method-discussion}

The effect of single-bounce reflection on RSS measurements, defined in Eq.~\eqref{eq:effect_of_reflection_db}, is a function of a single (reflection) point $\boldsymbol{p}$ on the object's boundary. Therefore, the single bounce reflection model does not require a precise geometrical model for the interacting object as long as the reflection coefficient $\Gamma$ is given, and the reflection point is parametrized using the excess path length $\Delta$ and transmitter $\boldsymbol{p}_t$ and receiver $\boldsymbol{p}_r$ positions.

Consider the effect of reflection $\zeta(\Delta)$ given in Eq.~\eqref{eq:effect_of_reflection_db}, and its spatial distribution depicted in Fig.~\ref{fig:reflection-2d}. In case the transmitter and the receiver are both stationary, the observed RSS variation is due to changes of excess path length $\Delta$ which can result from any movement of the interacting object. Let us denote the speed of the interacting object by $v \triangleq \| \boldsymbol{v} \|\ne {0}$ where $\boldsymbol{v}(t) \triangleq \frac{d}{dt} \boldsymbol{p}$ is the velocity of the object in the inertial frame of reference. The time derivative of $\Delta$ can be found by applying the chain rule of differentiation on Eq.~\eqref{eq:excess_path_length}, 
\begin{equation*}
	\begin{aligned}
	\frac{d}{d t}\Delta(t) = 
			\left[\frac{\boldsymbol{p} - \boldsymbol{p}_{r} }{\| \boldsymbol{p} - \boldsymbol{p}_{r} \| } + 
		          \frac{\boldsymbol{p} - \boldsymbol{p}_{t}}{\| \boldsymbol{p} - \boldsymbol{p}_{t} \| }\right]^T \boldsymbol{v}(t),
	\end{aligned}
\end{equation*}
where superscript $^T$ denotes matrix transpose, and the derivative exists only when $\boldsymbol{p} \ne \boldsymbol{p}_{r}$ and $\boldsymbol{p} \ne \boldsymbol{p}_{t}$. The bracketed term is the sum of two unity norm vectors. The norm of this vector approaches to $2$ when transmitter and receiver are close to each other compared to their distance to the point $\boldsymbol{p}$. As $\boldsymbol{p}$ gets closer to link-line, the time derivative of excess path length $\Delta$ becomes lower than the speed of the object so that it takes longer duration to observe valleys and peaks in RSS measurements. Conversely, in  the region far-away from link-line, small changes (on the order of a few cm) on the axis perpendicular to link-line shifts the RSS reading from detectable value down to noise floor as shown in Fig.~\ref{fig:reflection-2d}. Therefore, $\Delta_t$ must be selected close to a wavelength, e.g. $\Delta_t \approx 1.2 c_0 / f_c$.

The classifier assumes that the log-distance model parameters, path-loss exponent $\eta$ and average power measurement $\mathcal{P}_1$ at a reference distance $d_1$ are known. The path-loss exponent $\eta$ is site specific and needs to be determined for each deployment. Although one may measure $\mathcal{P}_1$ for each link, it may be advantageous to be able to estimate these values using the RSS measurements when the person is not in the environment (using LoS signal power estimates $\mathcal{P}_0$ of the links). If we suppose that the network nodes have similar hardware characteristics, one may assume $\mathcal{P}_1$ of different nodes is the same. In this case, only the frequency response of the radio RF front-end must be considered, and $\mathcal{P}_1(f_c) \equiv \mathcal{P}_1$ can be estimated along with $\eta$. Since the LoS signal power estimates of the links are linear with $\mathcal{P}_1(f_c)$ (cf. Eq.~\eqref{eq:log-distance-model}) and $d_{ij}$ are known for each transmitter and receiver, one can find the least square estimate $\mathcal{P}_1(f_c)$ for all $f_c$.

If the localization system acquires RSS measurements at different channels, the link level detector must make a decision for each of these. Similarly, the classifier must blacklist these links individually. Since the threshold values of the detectors do not depend on the communication frequency (cf. Eq.~\eqref{eq:zeta-envelopes}), the observations at different channels are compared with the same value. In this work, we simply assign a transmitter-receiver pair as not-detecting if majority of the channels do not detect presence of an interacting object. Similarly, we classify a transmitter-receiver pair as bad-link if majority of the channels are assigned in deep-fading. 

The distribution of $\hat{\boldsymbol{p}}$ in Eq.~\eqref{eq:position-estimate} is required for localization performance analysis. For this purpose, first note that the detector effectively is a binary quantizer, of which output has the discretized distribution of input. In other words, the detector output of any link between $i^{\text{th}}$ transmitter and $j^{\text{th}}$ receiver $\xi_{ij}$ in Eq.~\eqref{eq:probability-field} has the discretized distribution of $z_{ij}$. If the received signal experiences large number of reflections (scattering) from complex dielectric surfaces, the observations $z_{ij}$ are expected to have gamma distribution \cite{Abdi1999, Jakeman1982}. In addition, if the environment is cluttered such that the multipath signals experience multiples of reflection, scattering or diffraction, it is expected that the distribution of $z_{ij}$ approximates very closely to log-normal distribution \cite{Ohta1969} due to the statistical basis of log-normal received signal strength variation \cite{Coulson1998}. And second, note that the reconstruction method in Eq.~\eqref{eq:probability-field} requires to sum the scale values of the links indicating presence of an interacting object, which is a sum of random variables. If the number of links is large and distribution of $z_{ij}$ are identical (as it is expected when the environment is not cluttered), \emph{central limit theorem} can be invoked to conclude that the distribution of both axis of $\hat{\boldsymbol{p}}$ in Eq.~\eqref{eq:position-estimate} are Gaussian. On the other hand, when the environment is cluttered, both axis of $\hat{\boldsymbol{p}}$ have lognormal distribution since sum of log-normal random variables can be represented by a lognormal random variable \cite{Mehta2007}.

\begin{figure*}
\centering
\setlength{\tabcolsep}{1pt}
\begin{tabular}{C{0.25\textwidth}C{0.25\textwidth}C{0.25\textwidth}C{0.25\textwidth}}
	\subfloat[Experiment~\Rom{1}]{%
		\centering
		\renewcommand{\thesubfigure}{\roman{subfigure}}
		\begin{tabular}{r}
		\subfloat[]{\includegraphics[width=0.24\textwidth]{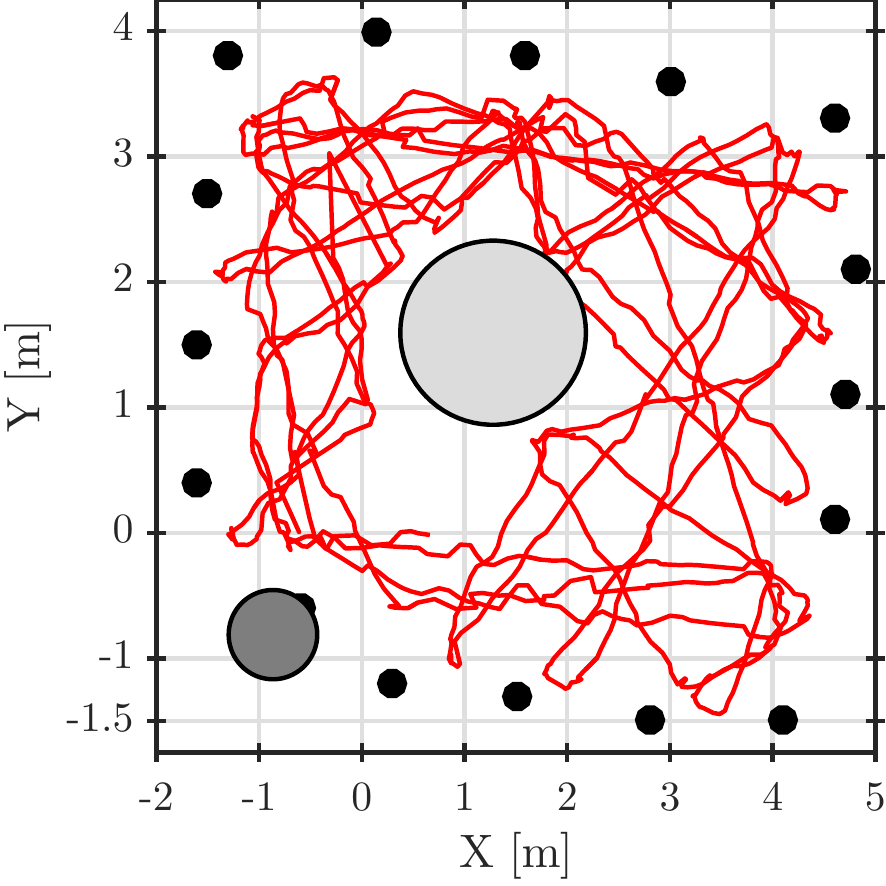}\setcounter{subfigure}{1}}\\
		\subfloat[]{\includegraphics[width=0.24\textwidth, height=3.2cm]{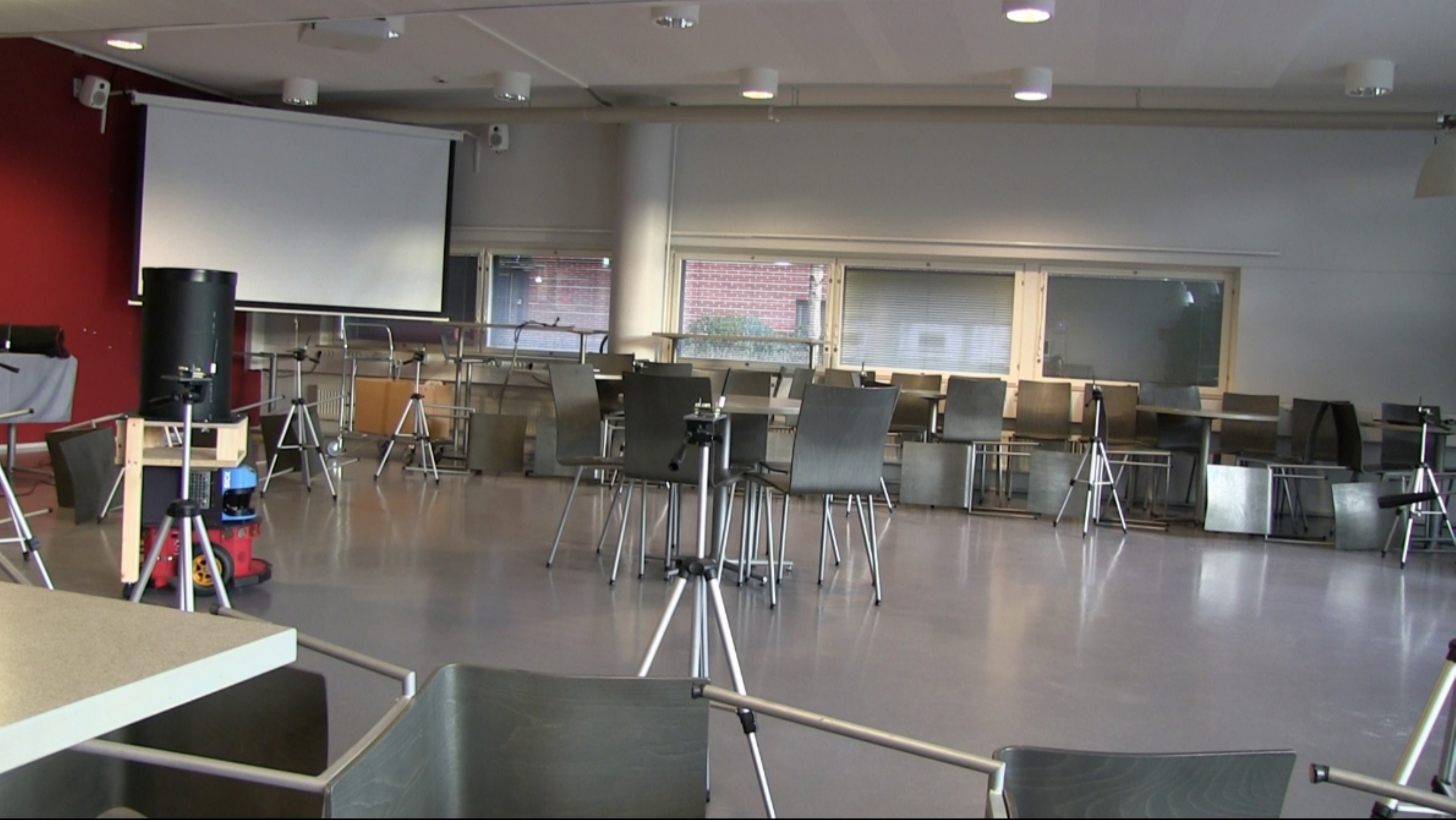}\setcounter{subfigure}{2}}
		\end{tabular}
		\setcounter{subfigure}{1}
		\label{fig:experiment1}
	}&
	\subfloat[Experiment~\Rom{2}]{%
		\centering
		\renewcommand{\thesubfigure}{\roman{subfigure}}
		\begin{tabular}{r}
		\subfloat[]{\includegraphics[width=0.24\textwidth]{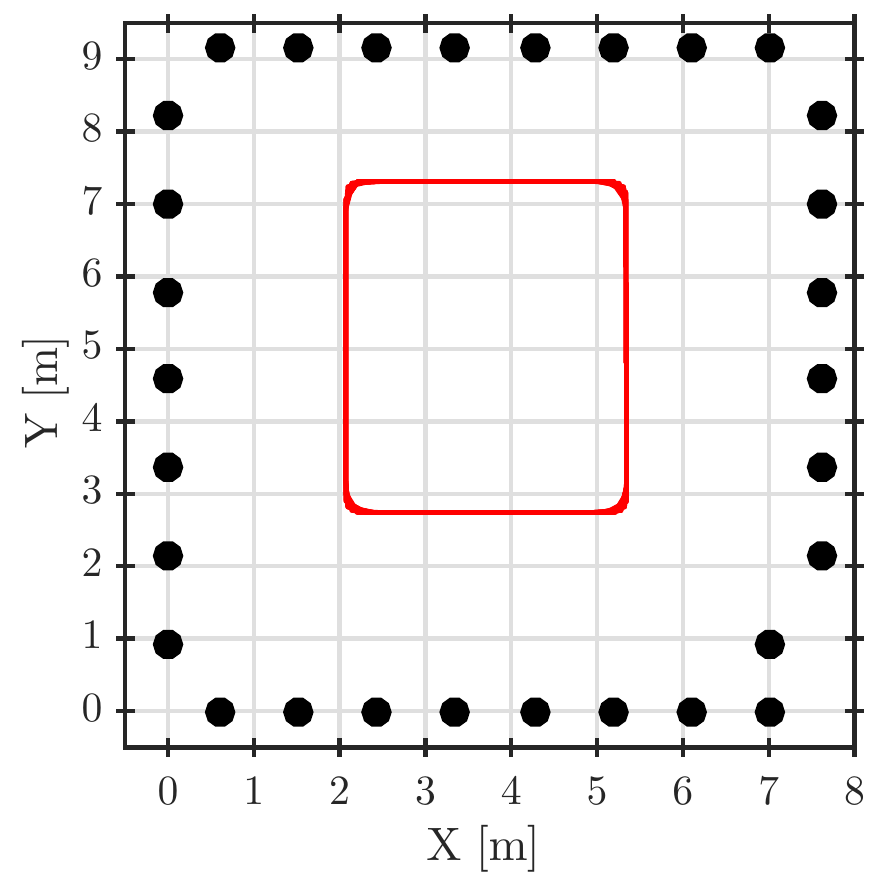}\setcounter{subfigure}{1}}\\
		\subfloat[]{\includegraphics[width=0.24\textwidth, height=3.2cm]{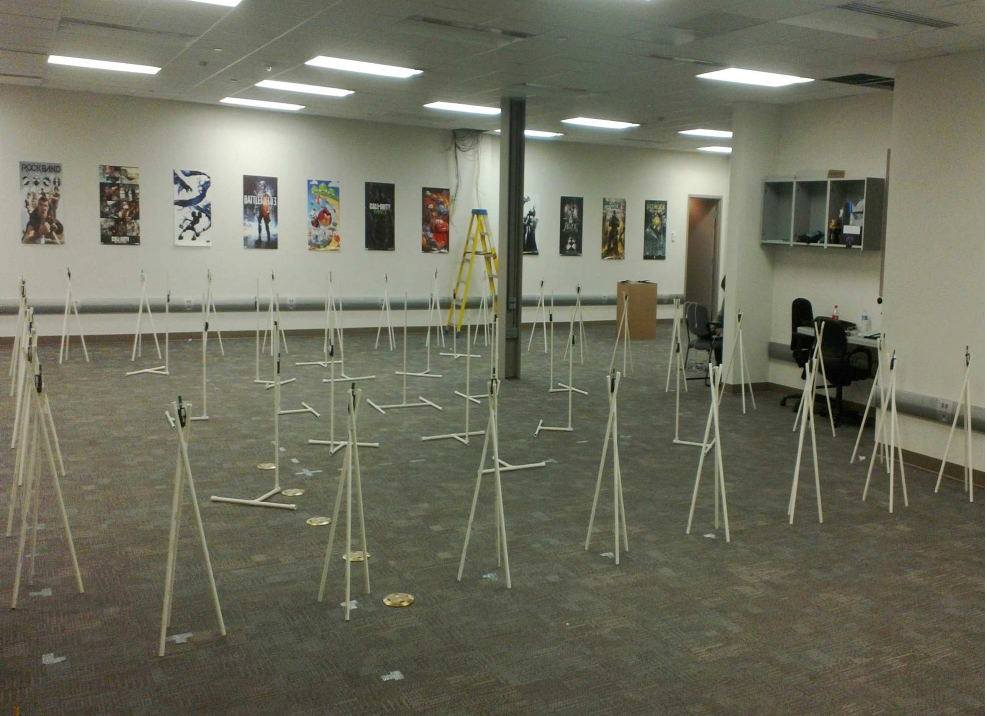}\setcounter{subfigure}{2}}
		\end{tabular}
		\setcounter{subfigure}{2}
		\label{fig:experiment2}
	}&
	\subfloat[Experiment~\Rom{3}]{%
		\centering
		\renewcommand{\thesubfigure}{\roman{subfigure}}
		\begin{tabular}{r}
		\subfloat[]{\includegraphics[width=0.24\textwidth]{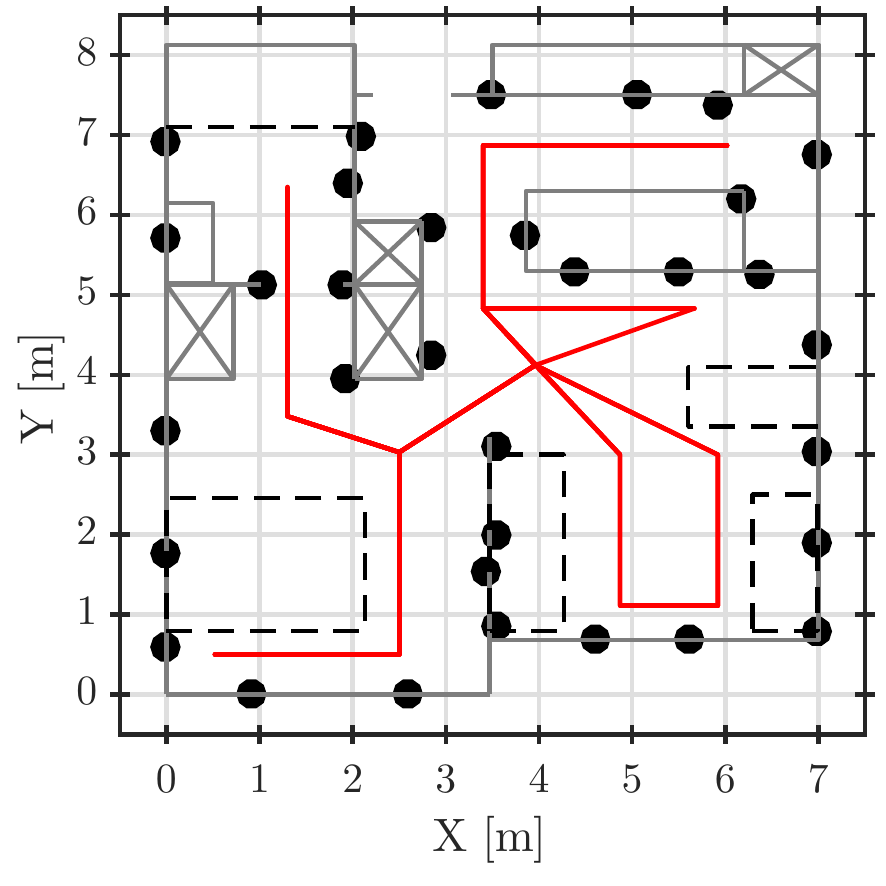}\setcounter{subfigure}{1}}\\
		\subfloat[]{\includegraphics[width=0.24\textwidth, height=3.2cm]{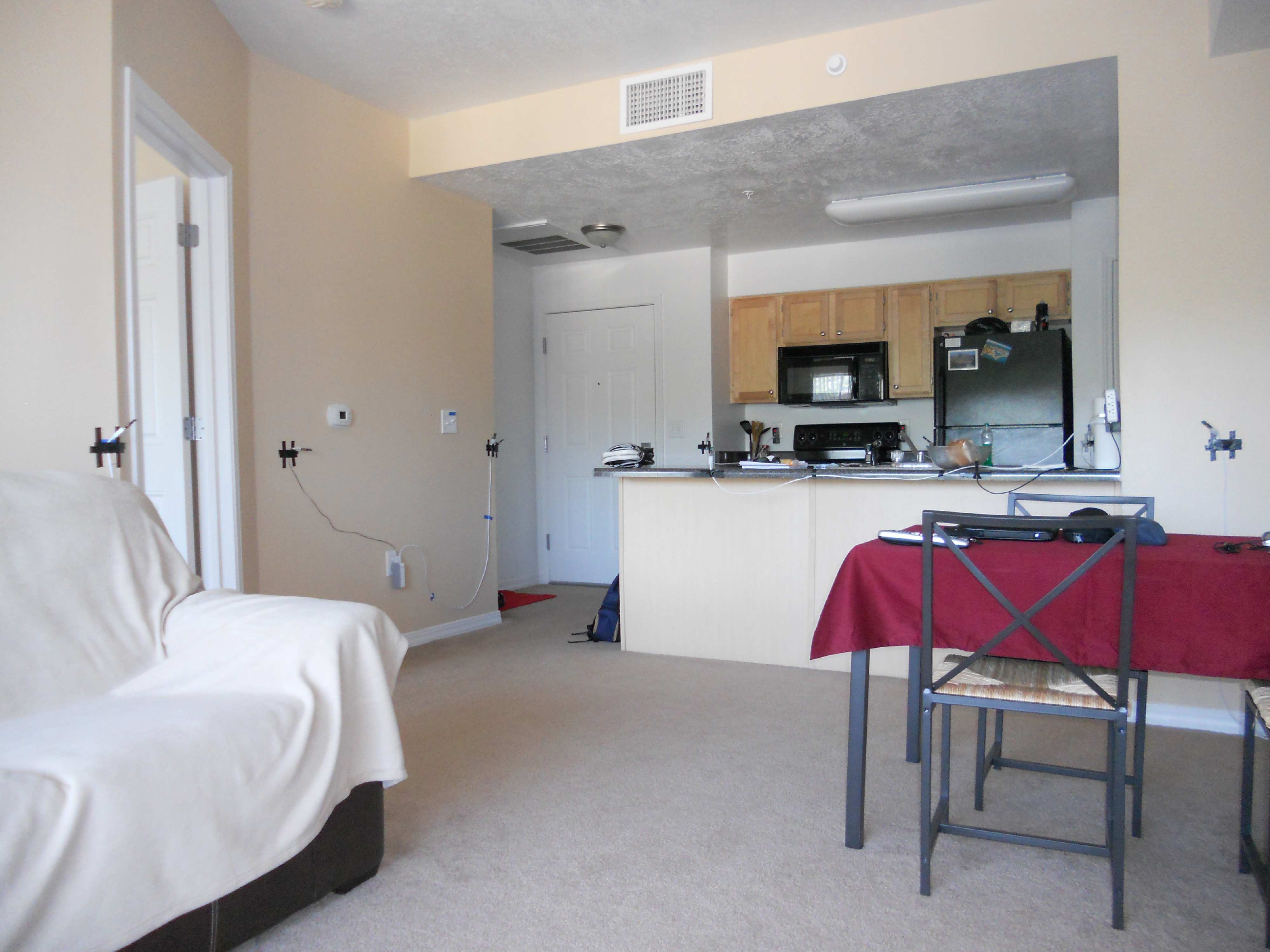}\setcounter{subfigure}{2}}
		\end{tabular}
		\setcounter{subfigure}{3}
		\label{fig:experiment3}
	}&
	\subfloat[Experiment~\Rom{4}]{%
		\centering
		\renewcommand{\thesubfigure}{\roman{subfigure}}
		\begin{tabular}{r}
		\subfloat[]{\includegraphics[width=0.24\textwidth]{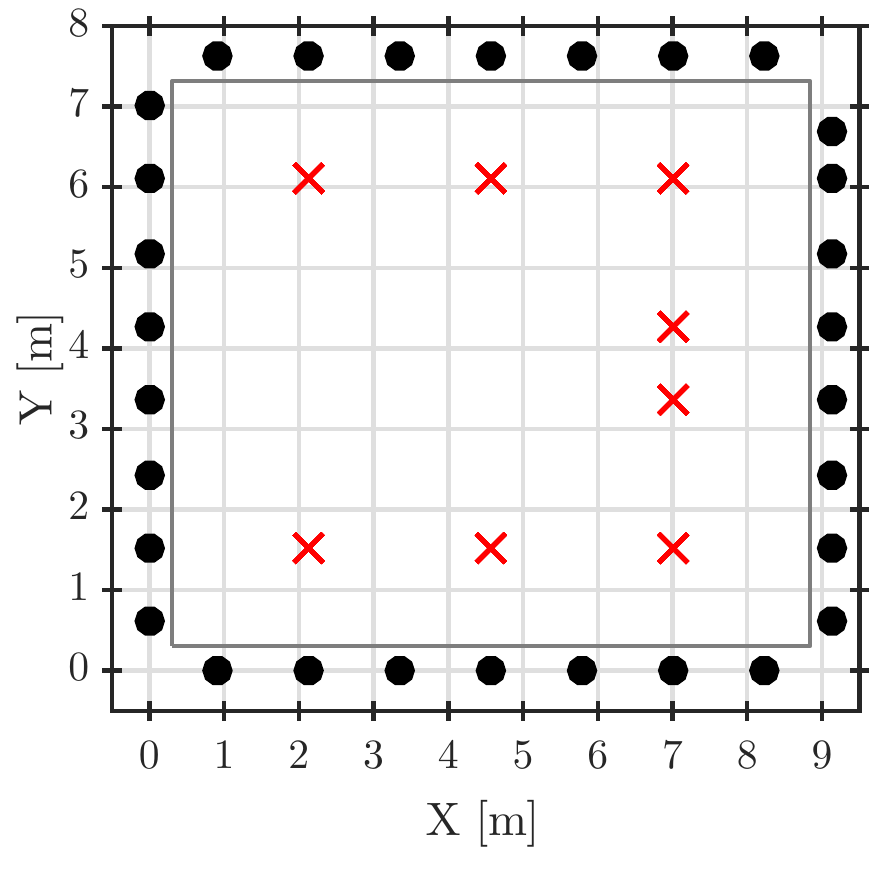}\setcounter{subfigure}{1}}\\
		\subfloat[]{\includegraphics[width=0.24\textwidth, height=3.2cm]{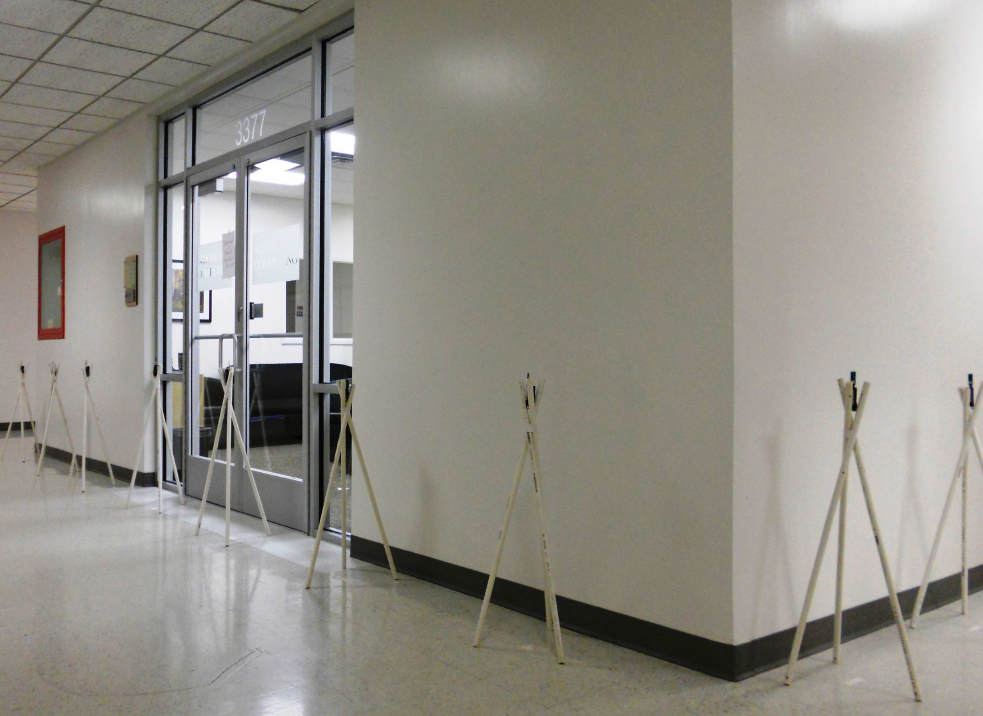}\setcounter{subfigure}{2}}
		\end{tabular}
		\setcounter{subfigure}{4}
		\label{fig:experiment4}
	}
\end{tabular}
\caption{Experiments~\Rom{1}--\Rom{4}: The interacting object either moves on the trajectory (shown as read line) or stands on predefined points (shown as red `x' markers) while the nodes (shown as black dots) are communicating. In (\rom{1}), the scene and the interacting object's trajectory. In (\rom{2}), deployment image}
\label{fig:experiments}
\end{figure*} 

\section{Experimental Evaluation}\label{sec:results}
In this section, empirical evaluation results of the system in Fig.~\ref{fig:system} is given. The experiment campaigns used for evaluations are described first, and then the evaluation criteria is defined. We discuss the findings after giving the results.


\subsection{Experiment Campaigns}\label{section:result-experiments}
In this work, we use four different experiments to validate the system.
The first experiment is performed using a mobile robot, which is equipped with a container that imitates electromagnetic properties of the human body at $2.4 \text{ GHz}$. The other experiments are conducted with human test subjects, but in different environments as described in the following and shown in Fig.~\ref{fig:experiments}. 


\subsubsection{Experiment~\Rom{1}}

A fully connected mesh network of $16$ IEEE~802.15.4 MAC/PHY standard compliant nodes is deployed in a region of $7 \times 6 \text{ m\textsuperscript{2}}$ as shown in Fig.~\ref{fig:experiment1}. The nodes have the hardware and software platform described in \cite{Yigitler2014}, and they are fixed on top of tripod stands approximately $85 \text{ cm}$ above the ground. The nodes have standard $2 \text{ dBi}$ gain omni-directional dipole antennas from Pulse Electronics. The operation of network is managed by the DFL management software running on a general purpose computer, which accesses the network through the sink node. The communication schedule, application software running on the nodes and the sink, and management software are as described in \cite{Yigitler2013}. During the experiments, the network is configured to transmit frames approximately every $5 \text{ ms}$ over channels $11 \text{, } 18 \text{ and } 26$ at $2.4 \text{ GHz}$ ISM band of of IEEE~802.15.4 MAC/PHY standard with $0 \text{ dBm}$ transmission power.

\begin{figure}[t]
\centering
\includegraphics[width=8cm]{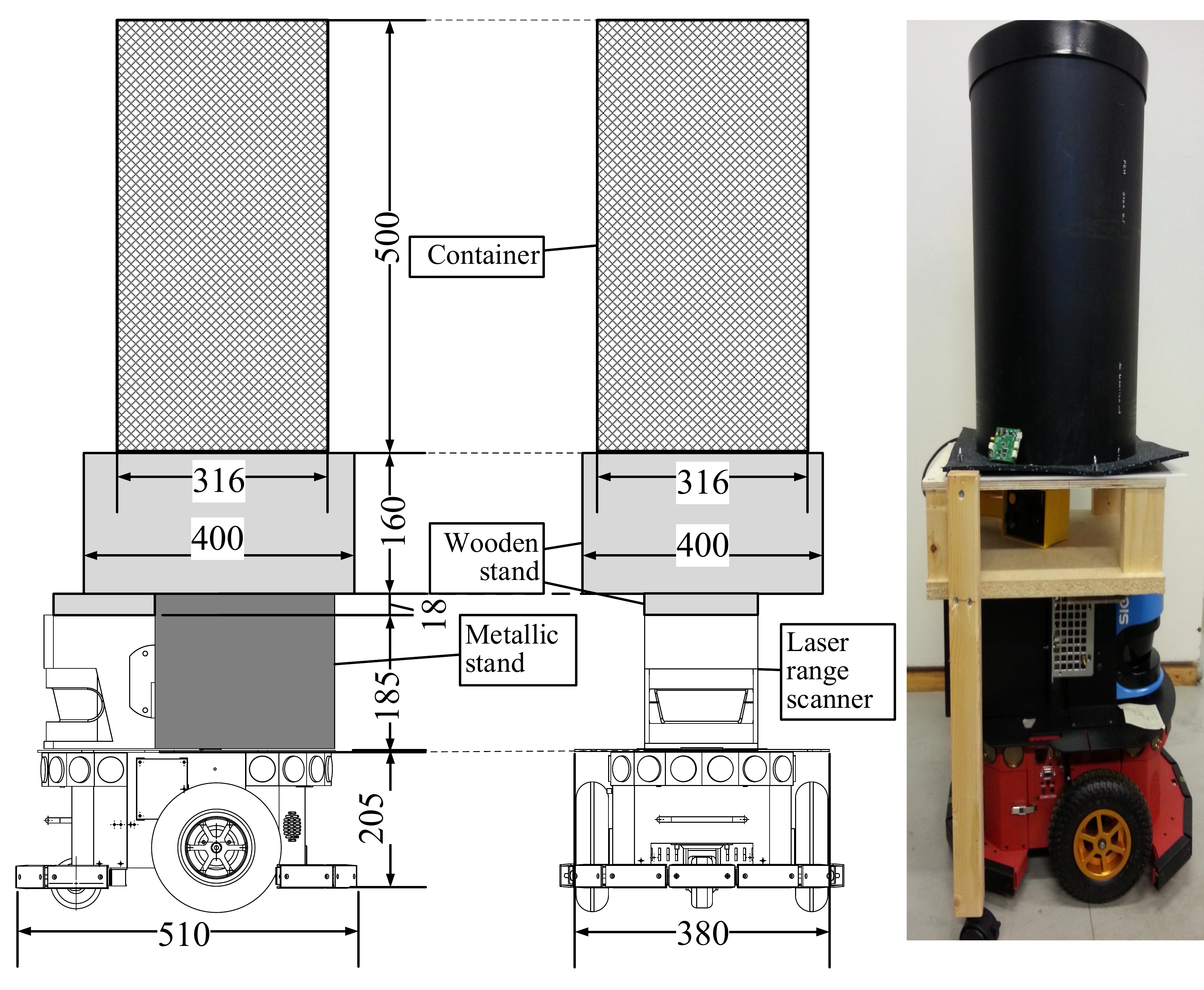}
\caption{Robot and human imitating container. All dimensions are in mm }
\label{fig:robot}
\end{figure}

In this experiment a mobile robot, which is Mobile Robots Pioneer 3 DX, is used as the test subject. A cylindrical container is placed on top of the robot as shown in Fig.~\ref{fig:robot}. The robot has total height of $1110 \text{ mm}$ above the ground and approximately mid-height of the container corresponds to nodes' antenna height. 
Essentially, the container is composed of two cylindrical pipes.
The outer pipe has $315 \text{ mm}$ diameter, $10 \text{ mm}$ thickness and is made of polyethylene. The inner pipe has $250 \text{ mm}$ diameter, $5 \text{ mm}$ thickness and it is made of PVC. The volume between the pipes is filled with body tissue simulating liquid manufactured by Speag \cite{tissueLiquid}. The liquid has the same electromagnetic properties as human tissue at $2.4 \text{ GHz}$. The container dimensions are selected based on the maximum payload weight of the robot.
%

The robot is equipped with inertial measurement unit, front and back sonar proximity sensors, and laser range scanner which are attached to an on-board computer. The computer estimates the robot's position using all of the aforementioned sensors and the environment map. The on-board computer has a dual-band wireless LAN adapter, which allows remote control over $5 \text{ GHz}$ link, and it is also attached to a sniffer node to synchronize robot's position logs to the network operation. The reader is referred to our recent work \cite{Yigitler2016} for possibilities enabled by this system.

During the experiment, the robot is commanded to navigate in a random fashion. It moves in a direction until it reaches to an obstacle, and it rotates until it can move forward without hitting to an obstacle. This way the robot traverses the region numerous times while recording its position. The robot sniffs the wireless communication in order to align the position logs to the network operation. This allows highly reliable RSS acquisition synchronized to very accurate robot position estimates. The robot's trajectory is shown in Fig.~\ref{fig:experiment1}(\rom{1}).

\subsubsection{Experiments~\Rom{2}--\Rom{4}}

In these experiments, fully connected mesh network of different number of IEEE~802.15.4 MAC/PHY standard compliant nodes are deployed in different environments. The nodes are Texas Instruments CC2531 USB dongles with $+4.5 \text{ dBm}$ maximum transmission power \cite{tidongle}, and they are placed on top of podiums at $100 \text{ cm}$ above the ground. The sensors communicate in round-robin fashion on multiple frequency channels. The software of different system components are as described in \cite{Bocca2013}.

During the experiments, the network is configured to transmit frames every $2.9 \text{ ms}$ over different channels at $2.4 \text{ GHz}$ ISM band of IEEE~802.15.4 MAC/PHY standard with $+4.5 \text{ dBm}$ transmission power. The person is asked to either stand on a specific position or move with constant speed on a predefined path. She follows placed markers inside the monitored areas and adjusts her walking pace with a metronome. In this way, each RSS measurement can be approximately associated to location of the person. These experiments are described in further detail in \cite{Kaltiokallio2013}.


{\textit{Experiment~\Rom{2}}}: $30$ nodes are deployed on the perimeter of $70 \text{ m\textsuperscript{2}}$ area as shown in Fig.~\ref{fig:experiment2}. The sensors are programmed to communicate over channels $11$, $17$, $22$, and $26$. During the test, the person enters the monitored area after the initial calibration period and walks along a predefined rectangular path as shown in Fig.~\ref{fig:experiment2}(\rom{1}), covering $14$ laps before leaving.

{\textit{Experiment~\Rom{3}}}: $33$ nodes are deployed in a single-floor, single-bedroom apartment of $58 \text{ m\textsuperscript{2}}$ as shown in Fig.~\ref{fig:experiment3}. The sensor podiums are not used; most of the sensors are attached on walls of the apartment and the remaining ones are placed elsewhere, e.g., on the edge of a marble counter in the kitchen or on the side of the refrigerator. 
The nodes are programmed to communicate over channels $15$, $20$, $25$, and $26$. During the test, a person is moving at constant speed along a path, shown in Fig.~\ref{fig:experiment3}(\rom{1}), several times. The path is chosen so as to cover the apartment.


{\textit{Experiment~\Rom{4}}}: In this experiment, the aim is to localize the person in a through-wall scenario with $30$ nodes that are deployed outside of the walls surrounding a lounge room of $70 \text{ m\textsuperscript{2}}$ as shown in Fig.~\ref{fig:experiment4}. The nodes are programmed to communicate over channels $11$, $15$, $18$, $21$, and $26$. During the experiment, the person is standing still at one of eight predefined positions as shown in Fig.~\ref{fig:experiment4}(\rom{1}). 

\subsection{Evaluation Metric}
Performance of the proposed method is evaluated with distance errors of the position estimates. Let us denote the true position of the interacting object at an arbitrary time instant by $\boldsymbol{p} \in \mathbb{R}^2$ and the position estimate with $\hat{\boldsymbol{p}}$. Then, the distance error is defined as
\begin{equation} \label{eq:distance-error}
	\varepsilon \triangleq \| \hat{\boldsymbol{p}} - \boldsymbol{p}\|.
\end{equation}

The distribution of $\varepsilon$ depends on the estimation error of each axis, whose distributions are discussed in Section~\ref{section:method-discussion}. In case they are normally distributed and if both axis have identical variances, then $\varepsilon$ has Rayleigh or Rice distribution depending on their means. If position error have different variance on each axis, then $\varepsilon$ has gamma distribution. If, on the other hand, the errors have log-normal distribution, $\varepsilon$ has log-normal distribution since product and sum of log-normal variables preserve the distribution. The robot has a regular and smooth surface, and interior of its container is relatively homogeneous so that $\varepsilon$ has Rayleigh distribution for Experiment~\Rom{1}. The human body, on the other hand, has irregular geometry, which causes different reflection coefficients for each orientation with respect to the incidence plane. Furthermore, it is composed of inhomogeneous dielectric boundaries and its geometry is time varying due to breathing. Thus, for the experiments with humans $\varepsilon$ is expected to have gamma distribution in case the environment is open (e.g. Experiment~\Rom{2}) or log-normal distribution if the environment is cluttered (e.g. Experiments~\Rom{3}). 

All three of the aforementioned distributions are skewed, that is to say, the first two moments of the distance error are not enough for quantitative performance evaluation. In this work, we evaluate the performance using the numerical values of first two moments and skewness (i.e. normalized third moment) of the distance error $\varepsilon$ defined in Eq.~\eqref{eq:distance-error} along with its histogram where possible.

\begin{table}[tb]
\renewcommand{\arraystretch}{1}
\renewcommand{\tabcolsep}{1 mm}
\caption{Experimental Evaluation Parameters} \label{table:parameters}
\centering
\begin{tabular}[c]{c c c L{4.7cm}}
\toprule [2pt]
\textbf{Symbol} & \textbf{Appearance}& \textbf{Value} & \multicolumn{1}{>{\centering\arraybackslash}m{47mm}}{\textbf{Explanation}}\\
\midrule[1pt] 
$\delta$ & Sec.~\ref{section:definitions} & $6.25$ & Pixel size in cm. \\
$\Gamma$ &Eq.~\eqref{eq:zeta-envelopes}& $0.50$ & Reflection coefficient for robot.\\
$\Gamma$ &Eq.~\eqref{eq:zeta-envelopes}& $0.35$ & Reflection coefficient for human.\\
$\eta$ &Eq.~\eqref{eq:zeta-envelopes}& $2.0$ & Path-loss exponent for open environments (Experiments~\Rom{1}--\Rom{3}).\\
$\eta$ &Eq.~\eqref{eq:zeta-envelopes}& $3.0$ & Path-loss exponent for through-wall scenario (Experiment~\Rom{4}).\\
$\mathcal{P}_d$ &Eq.~\eqref{eq:black-listing}& $-20$ & Fade level threshold.\\
$\Delta_{t}$ & Eq.~\eqref{eq:reflection-indicator} & $15.625$ & Maximum excess path length in cm. \\
$a$ & Eq.~\eqref{eq:probability-field2} & $0.75$ & Probability threshold scale. \\
\bottomrule [2pt]
\end{tabular}
\end{table}

\subsection{Results}

\begin{figure*}[thb]
\centering
\setlength{\tabcolsep}{1pt}
\begin{tabular}{C{0.33\textwidth}C{0.33\textwidth}C{0.33\textwidth}}
		\subfloat[]{\includegraphics[width=0.33\textwidth]{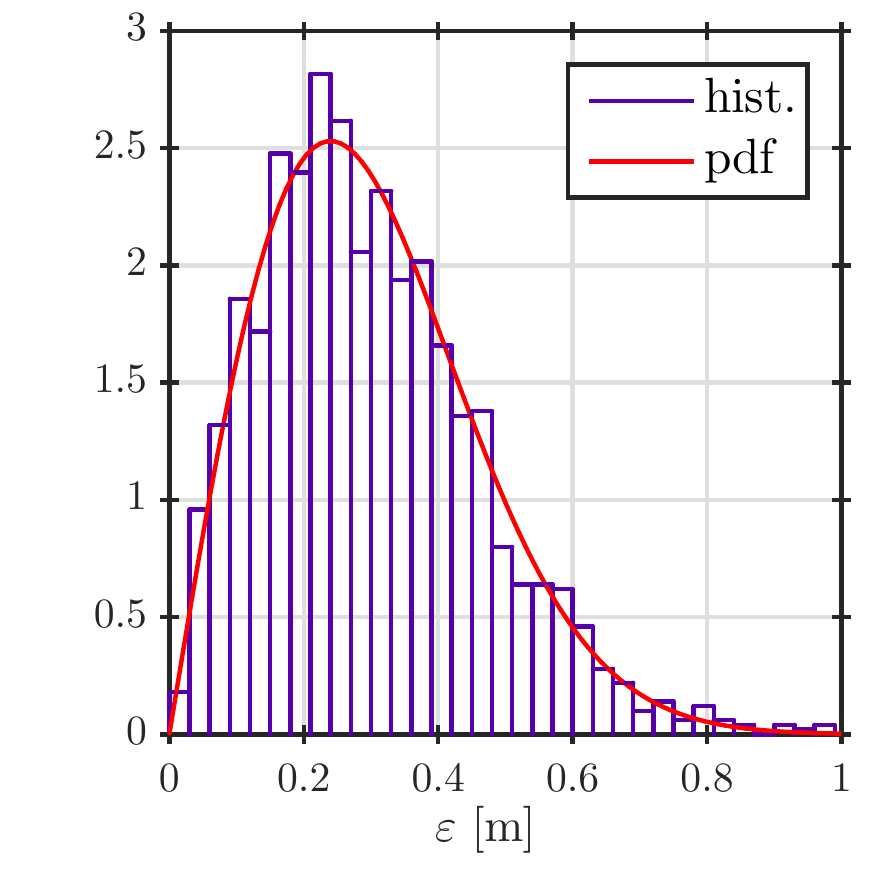}\label{fig:experiment1-density}}&
		\subfloat[]{\includegraphics[width=0.33\textwidth]{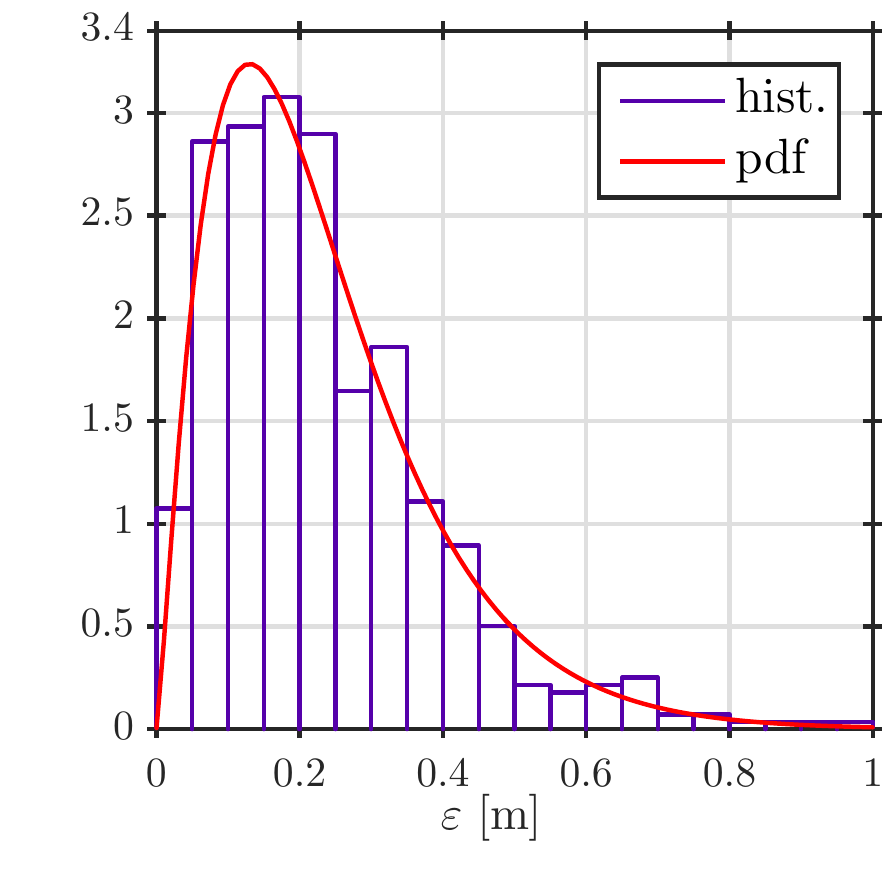}\label{fig:experiment2-density}}&
		\subfloat[]{\includegraphics[width=0.33\textwidth]{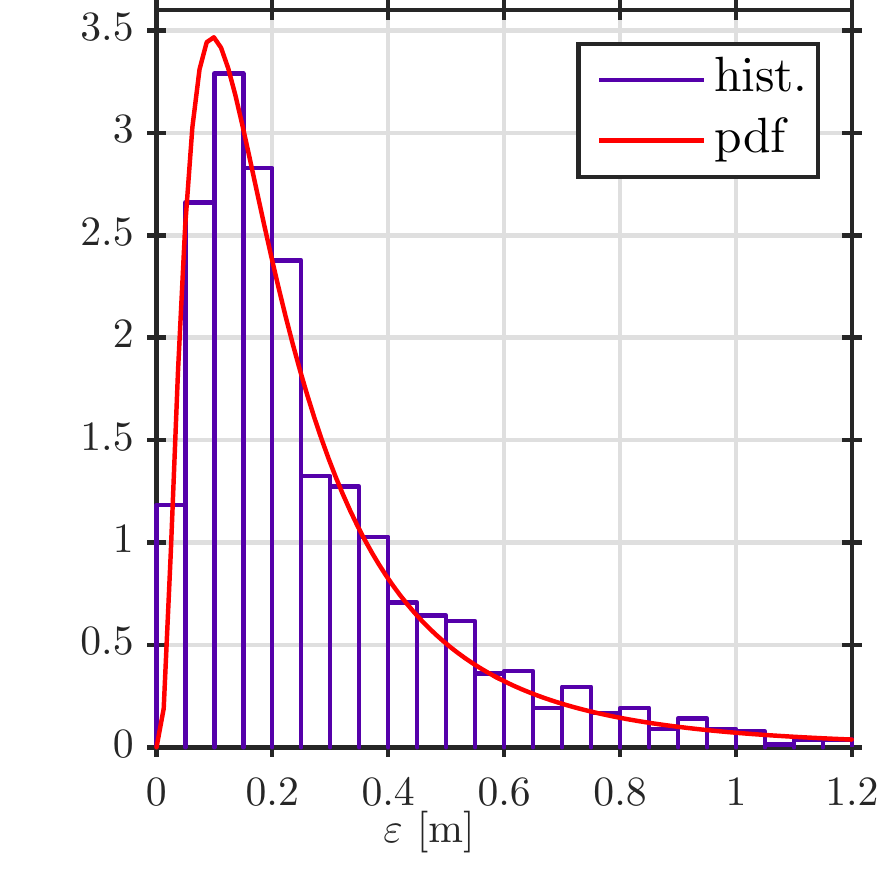}\label{fig:experiment3-density}}
\end{tabular}
	\caption{Distance error histograms and fitted distributions for Experiments~\Rom{1}--\Rom{3}. In (a), histogram and fitted Rayleigh density function for Experiment~\Rom{1}. In (b), histogram and fitted gamma density function for Experiment~\Rom{2}. In (c), histogram and fitted log-normal density function for Experiment~\Rom{3}.}
	\label{fig:experiment-densities} 
\end{figure*}

The method parameters are given in Table~\ref{table:parameters}. 

In the Experiments~\Rom{1}--\Rom{3}, the test subject moves to different locations so that a rich set of RSS measurements are acquired. In Experiment~\Rom{4}, on the contrary, the test subject stays still on eight predefined positions. For this experiment, the acquired data are only conclusive for assessing whether the method can be applied in through-wall scenarios.
Accordingly, in Table~\ref{table:error-stats}, the first three moments and distributions for Experiments~\Rom{1}--\Rom{3} are given, whereas for Experiment~\Rom{4} only the first three moments are tabulated. The results suggests that the proposed system can localize the person with high accuracy in terms of test subject's centroid position.

\begin{table}[tb]
\renewcommand{\arraystretch}{1.1}
\renewcommand{\tabcolsep}{1 mm}
\caption{Distance Error Statistics} \label{table:error-stats}
\centering
\begin{tabular}[c]{l c c c l}
\toprule [2pt]
 & \textbf{Exper.~\Rom{1} }& \textbf{Exper.~\Rom{2} }& \textbf{Exper.~\Rom{3}} & \textbf{Exper.~\Rom{4}} \\
\midrule[1pt] 
\textbf{Mean (m)} & $0.3085 $ & $0.2368 $ & $0.3096 $ & $0.4146 $ \\
\textbf{Variance (m\textsuperscript{2})} & $0.0388$ & $0.0291$ & $0.0980$ & $0.0874$ \\
\textbf{Skewness} & $2.6967$ & $2.1630$ & $2.5707$ & $3.1743$ \\
\midrule[1pt] 
\multicolumn{5}{c}{\textit{Kolmogorov-Smirnov Goodnes-of-Fit Test Results}}\\
\midrule[1pt]
\textbf{Distribution} & Rayleigh & Gamma & Log-normal & \\
\textbf{h-value} & $0$ & $0$ & $0$ & \\
\textbf{p-value} & $0.3845$ & $0.4697$ & $0.3237$ & \\
\bottomrule [2pt]
\end{tabular}
\end{table}

The fitted density functions and associated histograms of distance error $\varepsilon$ for Experiments~\Rom{1}--\Rom{3} are plotted in Fig.~\ref{fig:experiment-densities}. As shown, the histograms closely follows the fitted distributions, and their quantitative goodness-of-fit evaluation using Kolmogorov-Smirnov test with significance level of $0.05$, given on the lower part of Table~\ref{table:error-stats}, is in accordance with this statement. The \emph{h-value} signifies whether the null-hypothesis of empirical data does not belong to the tested distribution ($0$ indicates the data is drawn from the tested distribution). \emph{p-value} is the probability of observing a test statistic more extreme than the observed value under the null-hypothesis so that small values creates doubt about the \emph{h-value}. Therefore, distance error $\varepsilon$ has the tested distributions since the null-hypothesis is rejected for acceptable p-values.

\begin{table}[tb]
\renewcommand{\arraystretch}{1.1}
\renewcommand{\tabcolsep}{1 mm}
\caption{Mean Localization Error Comparison} \label{table:comparison}
\centering
\begin{tabular}[c]{l c c p{3.0cm} c}
\toprule [2pt]
 \textbf{Experiment} & \textbf{Mean (m)} & \textbf{Reported} & \multicolumn{1}{>{\centering\arraybackslash}m{30mm}}{\textbf{Method}} & \textbf{Ref.}\\
\midrule[1pt] 
Exper.~\Rom{1} & $0.3085$ & $0.2909$ & Network-Shadowing RTI &  \cite{Yigitler2016}\\
\midrule[0.5pt]
\multirow{2}{*}{Exper.~\Rom{2}} & \multirow{2}{*}{$0.2368$} & $0.1700$ & Fade-level RTI & \multirow{2}{*}{\cite{Kaltiokallio2013}}\\
 & & $0.2500$ & Channel-diversity RTI & \\
\midrule[0.5pt]
\multirow{2}{*}{Exper.~\Rom{3}} & \multirow{2}{*}{$0.3096$} & $0.2300$ & Fade-level RTI & \multirow{2}{*}{\cite{Kaltiokallio2013}}\\
 & & $0.2400$ & Channel-diversity RTI & \\
\midrule[0.5pt]
\multirow{2}{*}{Exper.~\Rom{4}} & \multirow{2}{*}{$0.4146$} & $0.3000$ & Fade-level RTI & \multirow{2}{*}{\cite{Kaltiokallio2013}}\\
 & & $0.7200$ & Channel-diversity RTI & \\
\bottomrule [2pt]
\end{tabular}
\end{table}

It must be noted that the method aims at detecting interacting object's presence using the reflected signal from its exterior, i.e., geometrical extend of the object is definitive for amount of distance error. However, previous works report their results as if the the test subject is a point, and their comparison is based the mean error. For comparative purposes, the mean distance errors for different RTI methods are given in Table~\ref{table:comparison} and compared with ones given in Table~\ref{table:error-stats}. The results in the table indicate that the proposed method performs comparable to the reported localization accuracy while providing clear advantages as discussed below.

\subsection{Discussion}\label{section:result-discussion}

The parameters in Table~\ref{table:parameters} indicates that the threshold values used by the classifier ($\mathcal{P}_d$), detector and reconstruction method ($\Delta_t$), and location estimator ($a$) are constant. The only parameters that may require adjustment are the model parameters. The reflection coefficient $\Gamma$ may require one to modify its value if dielectric boundary of the interacting object is significantly different (e.g. polyethylene for robot, clothing for human). For the human tracking purposes, its specified value ($\Gamma = 0.35$) is not expected to change. The path-loss exponent $\eta$ has its free-space value for open and obstructed environments ($\eta = 2$), whereas its value is increased for through-wall scenario ($\eta=3$). These values are in accordance with expected values as the number of walls between transmitter and receiver increases the path-loss exponent. Therefore, parameters of the system can be adjusted easily by considering the deployment. 

The method components can be distributed among the different network entities. Once the parameters are decided, a central node, which has deployment information, can calculate associated observation thresholds and distribute these to the network. The receivers can calculate LoS signal power for each neighbor, and classify them according to their fading levels using the calibration data acquired for some duration. This way, each receiver can store just the detector output by executing two simple instructions on the measured RSS: subtract the LoS signal power and compare it with the threshold. The central node is required to collect just binary valued assignments of each receiver, and calculate the spatial probability field using Eq.~\eqref{eq:probability-field}, which essentially is to sum the scale values. Therefore, the method is easy to implement and it can be distributed without data distribution management. 

For distributed implementations, the receivers are required to store only a single bit for each of their neighbors. This is a compressed representation of the information needed for localization purpose, which enables different possibilities. First, the nodes do not need to store complete measurements so that the memory requirement is reduced. Second, the information needs to be transmitted to the central node is reduced, which enables either faster measurement acquisition or improved scalability by just changing the transmission schedule but without complicated network management. Therefore, the proposed method enables new possibilities for RTI based DFL.

The traditional RTI require one to perform a matrix-vector product to form the image, which requires $LN$ floating point multiplications and $LN$ floating point additions, when there are $L$ links and $N$ pixels. Such operations require significant computational time on microcontrollers, which do not have a dedicated hardware for floating point arithmetic. For comparison, the proposed method require just $L^\prime N$ floating point additions when there are $L^\prime \le L$ links indicating interacting object's presence. Furthermore, in a typical DFL deployment, when the interacting object is on a specific position, only a subset of links can identify its presence so that most of the time we have $L^\prime \ll L$. Therefore, the method can be implemented on a microcontroller with limited computational capabilities.

The introduced system allows dividing the monitored area into sub-regions. The sub-regions can be selected so that the maximum number of pixels in effective regions of links does not exceed a maximum value so that a performance criteria can be satisfied. Another option is to divide the overall region into e.g. rooms, corridors, or any other area of interest. When there are a few number of people in the monitored area, most of these regions will not contribute to the overall localization effort since typically we have $L^\prime \ll L$. Suppose that there are $R$ regions of $N^\prime$ pixels, i.e. $N = R N^\prime$, and $R^\prime \ll R$ of these regions possess at least a pixel indicating occupied. As the monitored area increases, the number of regions $R$ increases. Since $R^\prime$ is expected to be bounded above due to limited area occupied by a person, the required number of floating point additions to form the occupancy field , which is $L^\prime R^\prime N^\prime$, is also bounded above. On the other hand, the traditional RTI also allows one to divide the region into sub-regions, if the system matrix is calculated and the image is constructed for each region separately. In other words, as the monitored area increases, required number of operations stays as  $L R N^\prime$ floating point multiplications and $L R N^\prime$ floating point additions. Therefore, the studied system improves scalability of the RTI systems greatly.

The computational requirements are low due to simplicity of the required mathematical operations. On the other hand, the memory storage requirement is dictated by the weight matrix $\boldsymbol{W}$. If one wishes to store it as a dense double precision matrix, it may consume significant amount of memory so that capacity of a constraint wireless sensor node would quickly be reached. This problem can be overcome by noting that the scale values in Eq.~\eqref{eq:reflection-scale} can be rewritten as
\begin{equation*}
\label{eq:reflection-scale2}
	\mathbbm{S}_{n,ij} = 
	\begin{cases}
	\frac{1}{\sum_{n}\mathbbm{1}_{n,ij}}\frac{1}{\sum_{ij}\{1\big/\sum_{n}\mathbbm{1}_{n,ij}\}} &\Delta_{n,ij} \le \Delta_t \\
	0 & \text{otherwise}
	\end{cases}
\end{equation*}
This equation implies that the complete weight matrix can be constructed using only the indicator matrix $\boldsymbol{\mathbbm{1}}$, which is a binary valued (and possibly sparse) matrix. 

The results presented in the previous section for four different experiments with different test subjects and in different environments allows us to conclude that the studied system shown in Fig.~\ref{fig:system} can be used for localization purposes. Furthermore, the mentioned properties imply that the proposed system is expected to improve the feasibility of RTI based DFL systems.

\section{Conclusion}\label{sec:conclusion}
In this paper, an imaging based device-free localization system aiming at reconstructing a spatial field of a person's presence is studied. Different than the available systems, the received signal strength variations is modeled using single-bounce reflections. The model is used for finding the decision thresholds of link-wise interacting object presence detectors. The detectors' output are input to a back-projection based reconstruction algorithm which uses pixel-wise occupancy assignments for each link. The interacting object is localized by finding the modes of the occupancy field. These methods significantly lower computational complexity of the localization effort, while using physically significant and repeatable parameters. The system is easy to implement and detectors can be distributed without data distribution management, while enabling compressing the localization information. The system is validated using four different measurement data. The results suggests that the localization accuracy is comparable to accuracy of the available methods.

\appendices
\section{Received Signal Power Measurement Noise} \label{appendix:noise-derivations}
Let us assume that $\hat{n}(t)$ in Eq.~\eqref{eq:received_signal} is a zero mean, complex white Gaussian noise process with two sided power spectral density of $N_0/2$. Let us further assume that power is calculated in digital domain using $K$ samples. In this case, the received signal power in Eq.~\eqref{eq:received_power} can be written as
\begin{equation}\label{eq:received_power_linear}
\mathrm{P}_r = (\alpha_1^2 + \alpha_2^2 + 2 \alpha_1\alpha_2\cos(\phi))\mathrm{P}_s + \frac{1}{K}\sum\limits_{k=1}^{K}{|\hat{n}_k|^2},
\end{equation}
where the noise process is assumed to be mean ergodic and independent of fading. The received signal power in logarithmic scale in Eq.~\eqref{eq:received_power} can be written as
\begin{equation*}\label{eq:simplified-received-signal-power}
	\mathcal{P}_r = 10 \log_{10}(\mathrm{P}_c) + n,
\end{equation*}
where $\mathrm{P}_c$ denotes all $\hat{n}_k$ independent terms in Eq.~\eqref{eq:received_power_linear}. Thus, the noise term is given by
\begin{equation}\label{eq:noise-n}
	n = {10}{\log_{10}} \left( 1 + \frac{1}{\mathrm{P}_c}{\frac{1}{K}\sum\limits_{k=1}^{K}{|\hat{n}_k|^2}}\right),
\end{equation}

Let us denote the sum by 
\begin{equation}\label{eq:sum-define}
	S_K = \frac{1}{K}\sum\limits_{k=1}^{K}{|\hat{n}_k|^2}.
\end{equation}
Since $|\hat{n}_k|^2$ is square sum of two independent zero-mean Gaussian random variables with equal variance of $\sigma^2$ within bandwidth of the communication system, it is a (central) Chi-Square random variable with two degrees of freedom \cite[pp.~45-46]{Proakis2008}, which is a Gamma random variable ($\sim\gamma(\vartheta, \theta)$) with shape parameters $\vartheta = 1$ and scale parameter $\theta = 2 \sigma^2$ \cite[ch.~17]{Johnson1994}. Then, $S_K$ is also a Gamma random variable $S_K\sim\gamma(K, 2\sigma^2 / K)$, since it is the sum of $K$ independent and identically distributed unity shape parameter Gamma random variables. Thus, the sum in Eq.~\eqref{eq:received_power_linear} is a Gamma random variable.

If the communication system is a direct sequence spread spectrum type, the number of samples acquired to calculate power measurement is expected to be very high. The distribution of $n$ for this case is defined by the asymptotic distribution of $S_K$ as $K \to \infty$, which can be found using stochastic convergence and limit theorems (see e.g. \cite[sec.~8.4]{Papoulis1991}). First, note that \emph{strong law of large numbers} implies that as $K \to \infty$ we have ${S_K} \to \mathrm{E}\left\{|\hat{n}_k|^2\right\} = 2 \sigma^2$ almost surely. Second, the \emph{central limit theorem} guarantees that distribution of the sum term ${S_K}$ approaches to Gaussian $\sim\mathcal{N}(2\sigma^2, 4 {\sigma^4}/{K})$. And finally, due to \emph{Berry-Esse\'en theorem}, the absolute point-wise difference between CDF of ${S_K}/{K}$ ($\mathbb{F}_{{S}_K}(\rho)$) and Gaussian CDF ($\mathbb{G}(\rho)$) satisfies \cite{Korolev2010}
\begin{equation}\label{eq:berry-esseen}
	\left| \mathbb{F}_{{S}_K}(\rho) - \mathbb{G}(\rho)\right| \le \frac{0.33554(\mu_3 + 0.415\mu_2\sqrt{\mu_2})}{\mu_2\sqrt{\mu_2}\sqrt{K}},
\end{equation}
where $\mu_r$ denotes $r^{\text{th}}$ central moment of $|\hat{n}_k|^2$ (for any $k \in[1,K]$). Since the non-central moments of a Gamma variate $\gamma(\vartheta, \theta)$ are given by \cite[p.~339]{Johnson1994}
\begin{equation}\label{eq:gamma-moment-r}
\mu_r^{\prime} = \theta^r \frac{\gamma(\vartheta + r)}{\gamma(\vartheta)},
\end{equation}
where $\gamma(\cdot)$ is the \emph{gamma} function, and $\mu_r^{\prime}$ is the $r^{\text{th}}$ non-central moment, it is a simple matter to show that the right hand side of Eq.~\eqref{eq:berry-esseen} is $0.8103291 / \sqrt{K}$. Therefore, for large $K$, $S_K$ is almost deterministic, and the degree of uncertainty is Gaussian distributed within confidence level specified by Eq.~\eqref{eq:berry-esseen}. 

Let us define $\bar{S}_K \triangleq S_K / \mathrm{P}_c$ for notational convenience while investigating the asymptotic behavior of $n$ defined in Eq.~\eqref{eq:noise-n}. Since the total received signal power $\mathrm{P}_c$ is a deterministic scalar, if ${S}_K$ is approximately Gaussian, $\bar{S}_K$ also approximates to Gaussian $\mathcal{N}(2\sigma^2/\mathrm{P}_c, 4\sigma^4/(K\mathrm{P}_c^2))$. 
Accordingly, noise $n$ in Eq.~\eqref{eq:noise-n} converges to 
\begin{equation}\label{eq:n-asymptotic-mean}
	 \lim\limits_{K \to \infty} n = \mathrm{E}\{n\} = 10\log_{10}\left( 1 + \frac{2}{\mathrm{P}_0 / \sigma^2}10^{-\frac{\zeta(\Delta)}{10}} \right),
\end{equation} 
where we have used the monotonicity of $\log(\cdot)$ function, and diminishing variance, $\mathrm{P}_c = \mathrm{P}_0 10^{\zeta(\Delta)/10}$, and ${\mathrm{P}_0}/{\sigma^2}$ is \emph{signal-to-noise ratio} (SNR) of the LoS signal. One consequence of Eq.~\eqref{eq:n-asymptotic-mean} is the impact of noise will diminish whenever the SNR of the LoS signal is very high. Since for a specific radio receiver $\sigma^2$ is fixed and equal to noise floor, the SNR variation is solely due to changes in received signal power, and for high SNR conditions the RSS statistics are dictated by quantization of the power measurements.

In case $K$ is very large, an accurate statistical model for RSS can be obtained by assuming Gaussian $\bar{S}_K \sim \mathcal{N}(2\sigma^2/\mathrm{P}_c, 4\sigma^4/(K\mathrm{P}_c^2))$, whose mean and variance both depend on received signal power $\mathrm{P}_c$. In this case, the distribution of $n$ can be found in closed form, and then the impact of quantization should be analyzed statistically \cite{Widrow1996}. Consequently, the quantized RSS follows the distribution of $n$ whose moments are modified according to Sheppard's corrections \cite[ch.~1]{Johnson1994}. 

%


\ifCLASSOPTIONcaptionsoff
  \newpage
\fi

\bibliographystyle{IEEEtran}

\end{document}